\documentclass[%
 reprint,
 amsmath,amssymb,
 aps,
pra,
]{revtex4-2}
   
\usepackage{graphicx}
\usepackage{dcolumn}
\usepackage{bm}
\usepackage[caption=false]{subfig}
\usepackage{xcolor}
\usepackage{physics}
\usepackage{comment}

\begin{document}

\preprint{APS/123-QED}

\title{Quantum ratchet with Lindblad rate equations}

\author{Luis Octavio Casta\~nos-Cervantes$^{\dagger}$}
\author{Jes\'{u}s Casado-Pascual}

\email{ luis.castanos@tec.mx, jcasado@us.es}

\affiliation{$^{\dagger}$Facultad de Ingenier\'{i}a, Universidad Nacional Aut\'{o}noma de M\'{e}xico, Circuito Escolar 04360, C.U., Coyoac\'an, 04510 Ciudad de M\'exico, M\'{e}xico \\
$^{\dagger}$Tecnol\'ogico de Monterrey, School of Engineering and Sciences, Ciudad de M\'{e}xico 14380, M\'{e}xico \\
$^{*}$F\'{i}sica Te\'{o}rica, Universidad de Sevilla, Apartado de Correos 1065, 41080 Sevilla, Spain}

\date{\today}

\begin{abstract}
A quantum random-walk model is established on a one-dimensional periodic lattice that fluctuates between two possible states. This model is defined by Lindblad rate equations that incorporate the transition rates between the two lattice states. Leveraging the system's symmetries, the particle velocity can be described using a finite set of equations, even though the state space is of infinite dimension. These equations yield an analytical expression for the velocity in the long-time limit, which is employed to analyze the characteristics of directed motion. Notably, the velocity can exhibit multiple inversions, and to achieve directed motion, distinct, nonzero transition rates between lattice states are required.
\end{abstract}

\maketitle

\section{Introduction}

Directed motion of particles in systems subjected to deterministic or stochastic unbiased driving forces has garnered significant and continuous attention~\cite{ReimannR, MolecularMotors, ArtificialBrownianMotors}. This phenomenon, commonly called \textit{ratchet effect}, has found important applications in physics~\cite{ReimannR}, chemistry~\cite{MolecularMotors}, biology~\cite{ArtificialBrownianMotors,India}, and nanotechnology~\cite{CommPhys}. The ratchet effect has not only been studied in systems composed of particles but has also been analyzed in extended systems~\cite{Marchesoni1996,*Salerno2002,*Salerno2002b,*MoralesMolina2003,*MoralesMolina2005,*SnchezRey2016,*CasadoPascual2019}.  From a theoretical perspective, significant efforts have been made to unravel the underlying mechanisms behind the emergence of directed motion~\cite{ReimannR,ArtificialBrownianMotors}. To make directed motion possible, it has been shown that specific spatiotemporal symmetry and supersymmetry conditions must be broken~\cite{Reimann2001}. Another intriguing phenomenon, frequently observed in these systems, is the reversal of current direction when a system parameter varies~\cite{Doering1994,Elston1996,Casado2006,Casado2018}. 

The ratchet effect, which was originally analyzed in classical systems, was later extended to the quantum domain~\cite{QR1997, Yukawa1997,India}. Specifically, research has demonstrated that quantum phenomena, such as tunneling and wave packet dispersion, can either enhance the ratchet current~\cite{SciRep2014} or lead to a reduction in transport efficiency~\cite{PRA842011}. A crucial factor contributing to ratchet behavior is the violation of time-reversal symmetry~\cite{PRB100,PRB2019phonon}. This can be accomplished, for example, through the irreversibility of dissipative effects. In this context, investigations have been conducted into the quantum dynamics of a particle in an asymmetric potential with Ohmic~\cite{PRB2019scaling} and super-Ohmic dissipation~\cite{PRB100}. It is also worth mentioning spin ratchets~\cite{SR1}, with a particular emphasis on dissipative spin ratchets~\cite{DSR1,DSR2}, where the violation of time-reversal symmetry is attributed to both dissipative effects and external magnetic fields. The ratchet effect has also been studied in the absence of dissipation where the system follows Hamiltonian dynamics (see, e.g., Ref.~\cite{Denisov2007}). In this case, the particle is typically subjected to potentials that break certain symmetries, as seen in the flashing ratchet potential considered in Ref.~\cite{Science2009}, the delta-kicked model in Ref.~\cite{QDK}, and the Bose-Hubbard model in Ref.~\cite{Pellegrini}. These types of potentials can be created, for example, using optical lattices~\cite{15,Science2009,QR2023}. Furthermore, these systems have also been used to study the relationship between many-body quantum chaos and entanglement in quantum ratchet systems~\cite{PRL1202018}, as well as the emergence of directed transport in interacting chaotic systems~\cite{sanku}.

Experimental investigations of quantum ratchets have been continually carried out since one of the original proposals involving a SQUID device~\cite{Zapata1996}. For instance, in Ref.~\cite{PhysicaE2018}, the observation of a polarization-sensitive magnetic quantum ratchet current was reported, where the direction and magnitude are determined by the orientation of the electric field. In another study~\cite{SciRep2014}, a scattering quantum ratchet was demonstrated, involving a directional flow of electrons in a two-dimensional electron gas. Furthermore, Ref.~\cite{NatureNano2013} presented an electronic quantum ratchet in graphene layers. In addition to these, quantum ratchets have also been realized in optical systems, as seen in the case of a delta-kicked photonic quantum ratchet~\cite{SciBull2015} and Bose-Einstein condensates in an optical lattice~\cite{QR2023}.

To gain a deeper understanding of the physics underlying the ratchet effect, it is advantageous to develop simplified models that retain the essential characteristics of the phenomenon under study and enable the extraction of analytical solutions~\cite{Casado2006, Polonica}. In particular, in Ref.~\cite{Casado2006} it is introduced a classical, one-dimensional random-walk model that allows the analytical study of the inversion of directed motion in a spatially symmetric fluctuating lattice. In this paper, we consider a quantum version of that model where some of the classical transitions between neighboring sites are replaced with coherent transitions. The objective of our study is to determine what different physical phenomena appear due to the quantum effects associated with the coherent transitions. Notably, it is found that multiple inversions of directed motion appear, a phenomenon that is absent in the classical case analyzed in Ref.~\cite{Casado2006}. The evolution of the system considered here is mathematically described using Lindblad rate equations~\cite{budini2006,Pellegrini}, which implies a non-Markovian evolution for the system under study. Our model allows us to derive an analytical expression for the long time limit of the particle's velocity. This holds particular appeal because non-Markovian systems are typically amenable only to numerical treatment. The analytical solution for the velocity enables us to examine the circumstances under which directed motion exists and multiple current reversals occur. To obtain the analytical expression for the velocity, the approach for classical master equations presented in Ref.~\cite{Derrida} is extended to Lindblad rate equations. Using this extension, the infinite number of equations describing the evolution of the matrix elements of the system density operator is reformulated in terms of a finite number of equations that can be solved analytically.

The structure of the remaining sections in this paper is as follows. In Sec.~\ref{Smodelo}, we introduce the model under investigation. The derivation of the Lindblad rate equations that describe the system's evolution is provided in Appendix~\ref{ApendiceDerivacion}. Section~\ref{Ssimetrias} examines the symmetries of the system, and in Sec.~\ref{SecV}, we establish the equations that describe the evolution of the velocity and its long-time behavior. Certain technical details required for deriving these equations can be found in Appendix~\ref{ApendiceV}. Section~\ref{Spropiedades} explores the key properties of the particle velocity in the long-time limit. Finally, in Sec.~\ref{conclusiones}, we present a summary and draw conclusions based on the aspects covered in this study.

\section{Description of the model}
\label{Smodelo}

\begin{figure}
   \centering
   \includegraphics[scale=0.45]{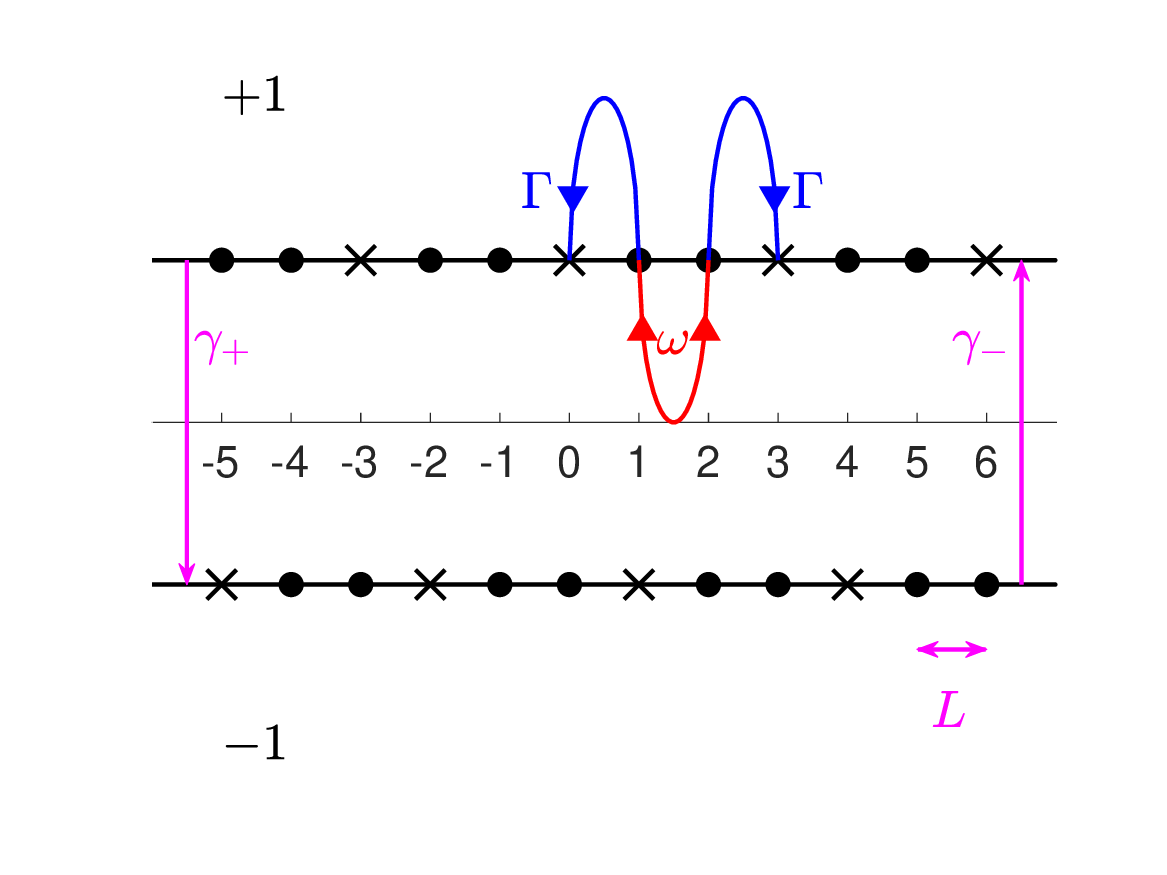}
   \caption{Sketch of the two possible states $\pm 1$ of the lattice under which the particle can move. The absorbing sites are depicted as crosses, while the nonabsorbing ones as solid circles. The horizontal axis appears in units of the distance $L$ between consecutive positions. The rate $\Gamma$ of the incoherent transitions, the frequency $\omega$ of the coherent transitions, and the rates $\gamma_{\pm}$ at which the lattice state changes are also illustrated.}
   \label{Figure1}
\end{figure}

Consider a quantum system composed of a particle that can be found at any site of the set $\mathcal{P} = \{ j L: j \in\mathbb{Z}\}$, with $L$ a given length and $\mathbb{Z}$ the set of integers (see Fig.~\ref{Figure1}). Let $\vert j \rangle$ be the quantum state associated with the position $jL$. These kets form an orthonormal basis $\left\{ \vert j \rangle: j\in \mathbb{Z} \right\}$ for the state space of the particle. The set $\mathcal{P}$ constitutes a one-dimensional lattice of sites that has two possible states denoted by $+1$ and $-1$. The lattice fluctuates randomly between these two states following a Markovian dichotomic process, with $\gamma_{\pm}$ being the transition rates from state $\pm 1$ to $\mp 1$. Consequently, the probability that the lattice sojourns in the state $\pm 1$ for a span of time $\tau$ and then changes to the state $\mp 1$ in the time interval between $\tau$ and $\tau+d\tau$ is given by $\gamma_{\pm} e^{-\gamma_{\pm} \tau} d\tau$ (see Ref.~\cite{Cox}). Therefore, the mean residence times at the lattice states $\pm 1$ are $T_{\pm} = \gamma_{\pm}^{-1}$.

As sketched in Fig.~\ref{Figure1}, in the lattice state $+1$ there are coherent transitions between the state vectors $\vert 3j+1 \rangle$ and $\vert 3j+2 \rangle$ at a frequency $\omega$. The corresponding sites $(3j+1)L$ and $(3j+2)L$ are called \textit{nonabsorbing}. In addition, there are incoherent transitions from $\vert 3j+1 \rangle$ and $\vert 3j+2 \rangle$ to $\vert 3j \rangle$ and $\vert 3j+3 \rangle$, respectively, at a rate $\Gamma$. No transitions are possible from state vectors of the form $\vert 3j \rangle$ to any other states and, for this reason, the sites $3jL$ are called \textit{absorbing}. In Fig.~\ref{Figure1}, the absorbing sites are depicted as crosses and the nonabsorbing as solid circles. The lattice state $-1$ is obtained by inverting the lattice state $+1$ with respect to any site of the form $(3j-1)L$. Alternatively, it can be obtained by translating the lattice state $+1$ by $(3j+1)L$ to the right or, equivalently, by $(3j+2)L$ to the left. Observe that both lattice states are periodic with period $3L$ and invariant under inversions about the absorbing sites. 

In terms of the jump operators $R_{j} = \vert j+1 \rangle\langle j \vert$, with $j\in\mathbb{Z}$, the above-described dynamics in each lattice state can be modeled by the superoperators
 \begin{eqnarray}
\label{LkLkp}
\mathcal{L}_{\pm}\boldsymbol{\cdot} &=& - \frac{i}{\hbar} \Big[ H_{\pm}, \boldsymbol{\cdot} \Big] + \mathcal{D}_{\pm}\boldsymbol{\cdot} , 
\end{eqnarray}
where $[\boldsymbol{\cdot} , \boldsymbol{\cdot} ]$ denotes the commutator, the Hamiltonians $H_{\pm}$ are given by
\begin{eqnarray}
\label{Bjk}
H_{\pm} &=& \hbar\omega \sum_{k\in\mathbb{Z}}  \Big( R_{3k\pm 1} + R_{3k\pm 1}^{\dagger}  \Big) , 
\end{eqnarray}
and the dissipators $\mathcal{D}_{\pm}$ are defined by
\begin{eqnarray}
\label{disipadorp}
 \mathcal{D}_{+}\boldsymbol{\cdot} &=& \Gamma \sum_{k\in\mathbb{Z}}\Bigg[ 
 R_{3k}^{\dagger} \boldsymbol{\cdot} R_{3k}  + R_{3k+2} \boldsymbol{\cdot} R_{3k+2}^{\dagger} \cr
 && \quad\quad\quad - \frac{1}{2}\Big\{ R_{3k} R_{3k}^{\dagger} + R_{3k+2}^{\dagger} R_{3k+2} , \boldsymbol{\cdot}  \Big\}  \Bigg] , \\
 \label{disipadorm}
\mathcal{D}_{-}\boldsymbol{\cdot} &=& \Gamma \sum_{k\in\mathbb{Z}}\Bigg[ 
 R_{3k} \boldsymbol{\cdot} R_{3k}^{\dagger}  + R_{3k+1}^{\dagger} \boldsymbol{\cdot} R_{3k+1} \cr
 && \quad\quad\quad - \frac{1}{2}\Big\{ R_{3k}^{\dagger} R_{3k} + R_{3k+1} R_{3k+1}^{\dagger} , \boldsymbol{\cdot}  \Big\}  \Bigg] ,
\end{eqnarray}
with $\{ \boldsymbol{\cdot}, \boldsymbol{\cdot} \}$ the anticommutator. The first term on the righthand side of Eq.~(\ref{LkLkp}) is responsible for the coherent evolution associated with the Hamiltonians $H_{\pm}$ in Eq.~(\ref{Bjk}). It describes coherent, \textit{tunnelinglike} transitions between those states associated with neighboring, nonabsorbing sites. By contrast, the second term, involving the dissipators $\mathcal{D}_{\pm}$ in Eqs.~(\ref{disipadorp}) and (\ref{disipadorm}), gives rise to incoherent transitions to the states associated with the absorbing sites. Notice that $\mathcal{L}_{\pm}$ have the form of a generator of a quantum dynamical semigroup~\cite{Breuer}. 

In analogy to \textit{composite stochastic processes}~\cite{vanKampen}, the density operator of the system, $\rho(t)$, can be expressed in terms of two positive operators $\rho_{\pm}(t)$ as $\rho (t) = \rho_{+}(t) + \rho_{-}(t)$. The evolution of these operators is governed by the \textit{Lindblad rate equations}
\begin{eqnarray}
\label{evolucion}
\frac{d}{dt} \rho_{\pm}(t) &=& \mathcal{L}_{\pm}  \rho_{\pm} (t) - \gamma_{\pm} \rho_{\pm}(t) + \gamma_{\mp} \rho_{\mp} (t) .
\end{eqnarray}
The concept of Lindblad rate equations was initially introduced in Ref.~\cite{budini2006} as a means to include non-Markovian effects in Lindblad-like master equations. Subsequently, it was further generalized in Ref.~\cite{Pellegrini}. To some extent, the Lindblad rate equations can be regarded as a quantum version of the composite stochastic processes introduced in Ref.~\cite{vanKampen}. In Appendix~\ref{ApendiceDerivacion}, we provide a derivation of Eq.~(\ref{evolucion}) that retains the spirit presented in Refs.~\cite{vanKampen,budini2006}.

The initial conditions for the Lindblad rate equations in Eq.~(\ref{evolucion}) depend on the probability $p_{\pm}(t_{0})$ that the lattice state is $\pm 1$ at the initial time $t_{0}$. Specifically, the initial conditions are $\rho_{+}(t_{0}) = p_{+}(t_{0})\rho(t_{0})$  and $\rho_{-}(t_{0}) = p_{-}(t_{0})\rho(t_{0})$ (see Appendix~\ref{ApendiceDerivacion}). 

\begin{figure}
	\centering
	\includegraphics[scale=0.45]{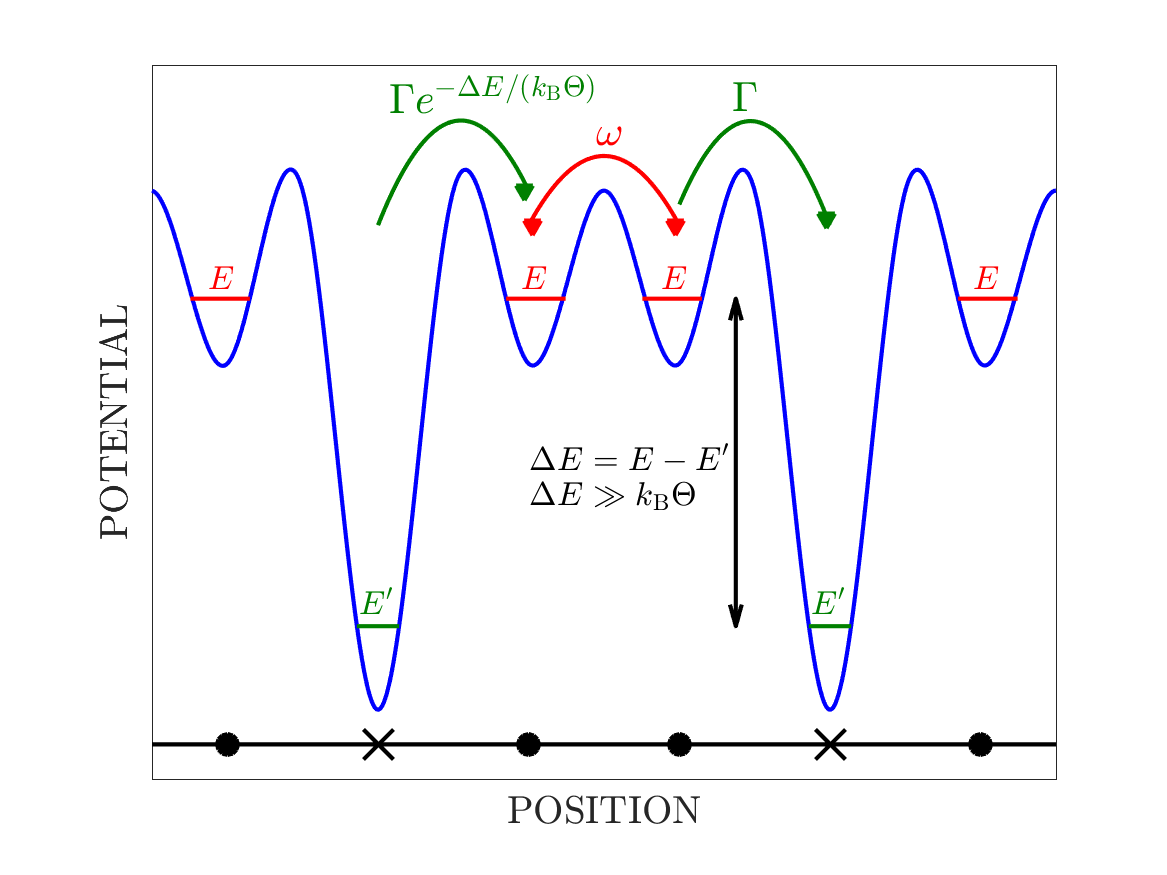}
	\caption{ Sketch illustrating a possible optical trapping potential for the experimental realization of the lattice states described in Fig.~\ref{Figure1}. The figure also depicts some relevant quantities characterizing the optical trapping potential. Specifically, $E$ and $E'$ represent the ground-state energies of the localized states around the nonabsorbing and absorbing sites, respectively, $\omega$ denotes the tunneling frequency between adjacent nonabsorbing sites, and $\Gamma$ indicates the incoherent transition rate from a nonabsorbing site to an absorbing one. The incoherent rate for the reverse transition is given by $\Gamma e^{-\Delta E/(k_{\mathrm{B}}\Theta)}$, where $\Delta E = E-E'$, $k_{\mathrm{B}}$ is the Boltzmann constant, and $\Theta$ is the absolute temperature of the thermal reservoir. If $k_{\mathrm{B}}\Theta \ll \Delta E$, the incoherent rate for the reverse transition becomes negligible compared to $\Gamma$.}
	\label{FigureNew}
\end{figure}

 A possible experimental implementation of the lattice states depicted in Fig.~\ref{Figure1} could be carried out using optical trapping potentials. In principle, quite arbitrary forms of trapping potentials can be generated by the fast-scanning of a focused laser beam~\cite{Optical1} or by holographic methods~\cite{Optical2}. Any of the lattice states considered in Fig.~\ref{Figure1} would correspond to a potential of the type depicted in Fig.~\ref{FigureNew}. As seen in this figure, the absorbing and nonabsorbing sites are represented by potential wells, with the depth of the absorbing wells much greater than that of the nonabsorbing ones. We assume that the system is in contact with a thermal reservoir at a sufficiently low absolute temperature $\Theta$. The ground-state energies of the localized states around the nonabsorbing and the absorbing sites are denoted by $E$ and $E'$, respectively. Transitions between adjacent nonabsorbing sites occur at a rate $\omega$ as a consequence of tunneling between the localized states in the corresponding wells. 
	
Given the significant disparity between the ground-state energies $E$ and $E'$, the transition between a nonabsorbing site to its adjacent absorbing site inevitably entails dissipation of energy, induced by the thermal reservoir. Unlike tunneling transitions, these incoherent transitions exhibit asymmetry~\cite{Breuer,QST}. Specifically, if $\Gamma$ represents the incoherent transition rate from a nonabsorbing site to an absorbing one, then the corresponding rate for the reverse transition is given by $\Gamma e^{-\Delta E/(k_{\mathrm{B}}\Theta)}$. Here, $\Delta E = E-E'$ is the energy difference between the localized ground states associated with the nonabsorbing and the absorbing sites (see Fig.~\ref{FigureNew}) and $k_{\mathrm{B}}$  the Boltzmann constant. As previously suggested,  the temperature $\Theta$ is assumed to be significantly lower than $\Delta E/k_{\mathrm{B}}$, rendering the incoherent transitions allowing departure from absorbing wells negligible, hence the term \textit{absorbing sites}.

The random transitions between the two lattice states shown in Fig.~\ref{Figure1} can be implemented by exploiting the possibility of generating optical trapping potentials dependent on an internal degree of freedom of the particle upon which they act~\cite{OL7}. In our scenario, this internal degree of freedom can be described by a spin with two possible states, denoted as $\ket{+}$ and $\ket{-}$. The optical trapping potentials corresponding to these two states, denoted as $U_+(x)$ and $U_-(x)$, respectively, would take shapes similar to that shown in Fig.~\ref{FigureNew}, with the potential $U_-(x)$ displaced a length $L$ to the right relative to $U_+(x)$. For an arbitrary spin state, the potential experienced by the particle would therefore take the form $U_+(x)\otimes \ketbra{+}{+}+U_-(x)\otimes \ketbra{-}{-}$. To induce transitions from the spin state $\ket{\pm}$ to $\ket{\mp}$, or equivalently from the optical trapping potential $U_{\pm}(x)$ to $U_{\mp}(x)$, it would suffice to apply $\pi$ pulses to the particle. In our case, the time interval $\tau$ elapsed between the application of two consecutive $\pi$ pulses should be a random variable with probability density $\gamma_{\pm} e^{-\gamma_{\pm} \tau}$, where the sign $\pm$ corresponds to the state taken by the spin during the interval under consideration.

\section{Symmetry considerations}
\label{Ssimetrias}

In this section, we discuss several symmetry properties of the superoperators $\mathcal{L}_{\pm}$ that arise as a consequence of the symmetries of the corresponding states of the lattice. In order to do this, for any $k\in\mathbb{Z}$, we introduce the \textit{translation operator} by $k$ units
\begin{eqnarray}
\label{Tk}
T_{k} = \sum_{j \in \mathbb{Z}} \vert j + k \rangle \langle j \vert , 
\end{eqnarray}
and the \textit{inversion operator} about site $kL$
\begin{eqnarray}
\label{Pik}
\Pi_{k} &=& \sum_{j\in\mathbb{Z}}\vert 2k -j \rangle\langle j \vert . 
\end{eqnarray}
From these definitions, it is clear that $\Pi_{k}$ is Hermitian and unitary and that $T_{k}$  is a unitary operator that satisfies $T_{j} T_{k} = T_{j + k}$ and $T_{k}^{\dagger} = T_{-k}$ for all $j,k\in\mathbb{Z}$.

Recall that both lattice states are spatially periodic with period $3L$ and also invariant under inversions about any of the absorbing sites (see Fig.~\ref{Figure1}). In addition, the lattice state $-1$ is obtained by translating the lattice state $+1$ by $(3j+1)L$ to the right or by inverting it with respect to any lattice sites of the form $(3j-1)L$ (see Fig.~\ref{Figure1}). The spatial periodicities of the lattice states $\pm 1$ are expressed by the invariance of $\mathcal{L}_{\pm}$ under the aforementioned translations 
\begin{equation}
\label{simetria1}
\mathcal{L}_{\pm} = \mathcal{T}_{3j} \mathcal{L}_{\pm} \mathcal{T}_{3j}^{\dagger} ,
\end{equation}
where we have introduced the \textit{translation superoperator} $\mathcal{T}_{k}\boldsymbol{\cdot} = T_{k} \boldsymbol{\cdot} T_{k}^{\dagger}$ and its adjoint with respect to the Hilbert-Schmidt inner product $\mathcal{T}_{k}^{\dagger}\boldsymbol{\cdot} = T_{k}^{\dagger}\boldsymbol{\cdot} T_{k}$. Analogously, the inversion invariances are expressed by
\begin{equation}
\label{simetria2a}
\mathcal{L}_{+} = \mathcal{P}_{3j} \mathcal{L}_{+} \mathcal{P}_{3j},
\end{equation}
and
\begin{equation}
\label{simetria2b}
\mathcal{L}_{-} = \mathcal{P}_{3j+1} \mathcal{L}_{-} \mathcal{P}_{3j+1} ,
\end{equation}
with the \textit{inversion superoperator} $\mathcal{P}_{k}\boldsymbol{\cdot} = \Pi_{k} \boldsymbol{\cdot} \Pi_{k}$. Finally, the aforementioned transformations from one lattice state to the other are represented by
\begin{equation}
\label{simetria3}
\mathcal{L}_{-} = \mathcal{T}_{3j+1} \mathcal{L}_{+} \mathcal{T}_{3j+1}^{\dagger}, 
\end{equation}
and
\begin{equation}
\label{simetria4}
\mathcal{L}_{\mp} = \mathcal{P}_{3j-1} \mathcal{L}_{\pm} \mathcal{P}_{3j-1} .
\end{equation}
Notice that, from the definitions of $\mathcal{T}_{k}$ and $\mathcal{P}_{k}$, it is clear that $\mathcal{T}_{k}^{\dagger}\mathcal{T}_{k} = \mathcal{T}_{k}\mathcal{T}_{k}^{\dagger} = \mathcal{I}$ and $\mathcal{P}_{k}^{2} = \mathcal{I}$, with $\mathcal{I}$ the identity superoperator.

A property that follows straightforwardly from Eqs.~(\ref{evolucion}) and (\ref{simetria4}) is that the operator $\tilde{\rho}_{\pm} (t) = \mathcal{P}_{-1} \rho_{\mp} (t)$ satisfies 
\begin{eqnarray}
\label{evolucion2}
\frac{d}{dt} \tilde{\rho}_{\pm}(t) &=& \mathcal{L}_{\pm}  \tilde{\rho}_{\pm} (t) - \gamma_{\mp} \tilde{\rho}_{\pm}(t) + \gamma_{\pm} \tilde{\rho}_{\mp} (t) .
\end{eqnarray}
Hence, $\tilde{\rho}_{+} (t)$ and $\tilde{\rho}_{-} (t)$ satisfy exactly the same equations as $\rho_{+}(t)$ and $\rho_{-}(t)$ but with $\gamma_{+}$ and $\gamma_{-}$ interchanged. The above described symmetries of $\mathcal{L}_{\pm}$ are used in the next sections to anticipate some properties of the system dynamics.

\section{Derivation of the particle velocity}
\label{SecV}

We are interested in characterizing the mean velocity of the particle. In order to do so, we first introduce the \textit{position operator}
\begin{eqnarray}
\label{Xp}
X &=& L \sum_{j\in \mathbb{Z}} j \vert j \rangle\langle j \vert ,
\end{eqnarray}
whose mean value at time $t$ is given by
\begin{eqnarray}
\label{X}
\langle X \rangle(t) &=& \mbox{Tr} \left[ X \rho(t) \right]  = \sum_{\alpha = \pm} \mbox{Tr} \left[ X\rho_{\alpha}(t) \right]  , \quad
\end{eqnarray}
where $\mathrm{Tr}(\boldsymbol{\cdot})$ denotes the trace. The \textit{mean velocity} is then defined as
\begin{eqnarray}
\label{vd}
v(t) &=& \frac{d}{dt} \langle X \rangle (t) .
\end{eqnarray}

Before establishing a method to calculate $v(t)$, let us discuss some consequences of the symmetries described in the previous section. As mentioned before, the operators $\tilde{\rho}_{\pm}(t)$ satisfy the same equations as $\rho_{\pm}(t)$ but with the rate parameters $\gamma_{+}$ and $\gamma_{-}$ interchanged [see Eqs.~(\ref{evolucion}) and (\ref{evolucion2})]. If the initial conditions are invariant under inversions about site $-L$, i.e., $\mathcal{P}_{-1}\rho_{\pm}(0) = \rho_{\pm}(0)$, it then follows from the uniqueness of the solution that $\tilde{\rho}_{\pm}(t;\gamma_{-},\gamma_{+}) = \rho_{\pm}(t;\gamma_{+},\gamma_{-})$, where the dependence on the rate parameters $\gamma_{\pm}$ has been written explicitly. Consequently, from the definition in Eq.~(\ref{X}) and the fact that $\mathcal{P}_{-1}X = -X-2LI$, with $I$ the identity operator, it follows that  $\langle X \rangle (t;\gamma_{+},\gamma_{-}) = -\langle X \rangle (t;\gamma_{-},\gamma_{+}) -2L$ and, therefore, $v(t;\gamma_{+}, \gamma_{-}) = -v(t;\gamma_{-}, \gamma_{+})$. Below it is shown that the long-time limit of the velocity $v_{\infty} = \lim_{t\rightarrow + \infty} v(t)$ is independent of the initial conditions. Thus, $v_{\infty}(\gamma_{+},\gamma_{-}) = - v_{\infty}(\gamma_{-},\gamma_{+})$ regardless of the initial conditions and, in particular, $v_{\infty} = 0$ if $\gamma_{+} = \gamma_{-}$. As a result, a necessary condition to have directed motion is that $\gamma_{+} \not= \gamma_{-}$.

Inspired by the approach of Ref.~\cite{Derrida}, in order to obtain an explicit expression for $v(t)$, it is convenient to introduce the operators
\begin{eqnarray}
\label{19}
\Lambda_{\pm}(t) &=& \sum_{j\in\mathbb{Z}}  \mathcal{T}_{3j}\rho_{\pm}(t)  ,
\end{eqnarray}
which are clearly invariant under translations by $3m$ units, i.e., $\mathcal{T}_{3m}\Lambda_{\pm}(t) = \Lambda_{\pm}(t)$. In addition, taking into account Eq.~(\ref{simetria1}), it readily follows that $\Lambda_{\pm}(t)$ satisfy the same equations as $\rho_{\pm}(t)$, i.e., Eq.~(\ref{evolucion}). In terms of the matrix elements
\begin{eqnarray}
\label{NP}
\lambda_{j,k}^{(\pm)}(t) &=& \langle j \vert \Lambda_{\pm}(t) \vert k \rangle, 
\end{eqnarray}
the invariance of $\Lambda_{\pm}(t)$ under translations by $3m$ units takes the form
\begin{equation}
\label{periodicidad}
\lambda_{j,k}^{(\pm)}(t) = \lambda_{j+3m,k+3m}^{(\pm)}(t) .  
\end{equation}
It should also be noticed that $\sum_{\alpha = \pm} \sum_{j=1}^{3} \lambda_{j,j}^{(\alpha)}(t) = \sum_{\alpha = \pm}\mbox{Tr}[\rho_{\alpha}(t)] = \mbox{Tr}[\rho(t)] = 1$.

Using the evolution equations for $\rho_{\pm}(t)$ in Eq.~(\ref{evolucion}), in combination with the properties of $\Lambda_{\pm}(t)$ discussed above, it can be verified that
\begin{eqnarray}
\label{v}
\frac{v(t)}{L} &=& \Gamma \Big[  \lambda_{3,3}^{(-)}(t)  - \lambda_{2,2}^{(-)}(t)  + \lambda_{2,2}^{(+)}(t) - \lambda_{1,1}^{(+)}(t) \Big] \cr
&& - 2\omega \mathrm{Im}\Big[ \lambda_{3,2}^{(-)}(t) + \lambda_{2,1}^{(+)}(t) \Big] , 
\end{eqnarray}
with $\mathrm{Im}(z)$ the imaginary part of the complex number $z$. For a detailed derivation of Eq.~(\ref{v}), see Appendix~\ref{ApendiceV}.

The quantum nature of our model is clearly evidenced in Eq.~(\ref{v}). Within this equation, two distinct terms of different nature are present. Firstly, the term $v_{\mathrm{q}}(t)/L = \smash{- 2\omega \mathrm{Im}[ \lambda_{3,2}^{(-)}(t) + \lambda_{2,1}^{(+)}(t)]}$ depends on the nondiagonal elements of the \textit{partial density operators} $\rho_{\pm}(t)$, commonly referred to as coherences.
These coherences lack a classical analog and are inherently quantum. In addition, the term $v_{\mathrm{c}}(t)/L = \smash{\Gamma [  \lambda_{3,3}^{(-)}(t)  - \lambda_{2,2}^{(-)}(t)}$ $\smash{ + \lambda_{2,2}^{(+)}(t) - \lambda_{1,1}^{(+)}(t)]}$ depends only on the diagonal elements of $\rho_{\pm}(t)$,  commonly referred to as populations. This term can be easily interpreted through classical arguments of population balance: the particle advances (retreats) as a portion of the population from sites immediately to the left (right) of the absorbing sites is absorbed by them.

To determine the average velocity in Eq.~(\ref{v}), it is first necessary to solve the evolution equations for the matrix elements of $\Lambda_{\pm}(t)$. At first sight, this would involve a system of an infinite number of coupled differential equations. Nevertheless, using the periodicity property in Eq.~(\ref{periodicidad}), this infinite system can be reduced to a finite number of differential equations for the matrix elements $\lambda_{j,k}^{(\pm)}(t)$ with $j,k\in\{1,2,3\}$, i.e., $18$ equations. The explicit form of twelve of these equations is
\begin{eqnarray}
\label{sistemaEc}
D_{-,0} \lambda_{1,1}^{(-)}(t)
&=&  \Gamma \Big[ \lambda_{3,3}^{(-)}(t) + \lambda_{2,2}^{(-)}(t) \Big] + \gamma_{+} \lambda_{1,1}^{(+)}(t) , \cr
D_{-,2} \lambda_{2,2}^{(-)}(t)
&=& i\omega \Big[ \lambda_{2,3}^{(-)}(t)  - \lambda_{3,2}^{(-)}(t) \Big]  + \gamma_{+} \lambda_{2,2}^{(+)}(t) , \cr
D_{-,2} \lambda_{3,3}^{(-)}(t)
&=& -i\omega \Big[ \lambda_{2,3}^{(-)}(t) - \lambda_{3,2}^{(-)}(t) \Big]  + \gamma_{+} \lambda_{3,3}^{(+)}(t) , \cr
D_{+,2} \lambda_{1,1}^{(+)}(t) 
&=& -i\omega \Big[ \lambda_{2,1}^{(+)}(t)  - \lambda_{1,2}^{(+)}(t) \Big] + \gamma_{-} \lambda_{1,1}^{(-)}(t) , \cr
D_{+,2} \lambda_{2,2}^{(+)}(t) 
&=& i\omega \Big[ \lambda_{2,1}^{(+)}(t) - \lambda_{1,2}^{(+)}(t) \Big]  + \gamma_{-} \lambda_{2,2}^{(-)}(t) , \cr
D_{+,0} \lambda_{3,3}^{(+)}(t)
&=&  \Gamma \Big[ \lambda_{2,2}^{(+)}(t) + \lambda_{1,1}^{(+)}(t) \Big] + \gamma_{-} \lambda_{3,3}^{(-)}(t) , \cr
D_{-,1} \lambda_{1,2}^{(-)}(t) 
&=& i\omega \lambda_{1,3}^{(-)}(t) + \gamma_{+} \lambda_{1,2}^{(+)}(t) , \cr
D_{-,1} \lambda_{1,3}^{(-)}(t)
&=& i\omega \lambda_{1,2}^{(-)}(t)  + \gamma_{+} \lambda_{1,3}^{(+)}(t) , \cr
D_{-,2} \lambda_{2,3}^{(-)}(t)
&=& -i\omega \Big[ \lambda_{3,3}^{(-)}(t) - \lambda_{2,2}^{(-)}(t) \Big]  + \gamma_{+} \lambda_{2,3}^{(+)}(t)  , \cr
D_{+,2} \lambda_{1,2}^{(+)}(t) 
&=& -i\omega \Big[ \lambda_{2,2}^{(+)}(t) - \lambda_{1,1}^{(+)}(t) \Big]  + \gamma_{-} \lambda_{1,2}^{(-)}(t) , \cr
D_{+,1} \lambda_{1,3}^{(+)}(t)
&=& -i\omega \lambda_{2,3}^{(+)}(t) + \gamma_{-} \lambda_{1,3}^{(-)}(t)  , \cr
D_{+,1} \lambda_{2,3}^{(+)}(t)
&=& -i\omega \lambda_{1,3}^{(+)}(t) + \gamma_{-} \lambda_{2,3}^{(-)}(t)  ,
\end{eqnarray}
where we have introduced the differential operators
\[
D_{\pm , j} = \frac{d}{dt} + \left( j\frac{\Gamma}{2} + \gamma_{\pm} \right) , 
\]
for $j=0,1,2$. The remaining six equations can be obtained by taking the complex conjugate of the last six equations appearing in Eq.~(\ref{sistemaEc}) and using that $\Lambda_{\pm}(t)$ are Hermitian. 

Taking into account the definition of $\Lambda_{\pm}(t)$, the initial conditions for the system of differential equations in Eq.~(\ref{sistemaEc}) can be obtained from the initial conditions $\rho_{\pm}(0)$ by $\lambda_{j,k}^{(\pm)}(0) = \sum_{m\in\mathbb{Z}} \langle j + 3m \vert  \rho_{\pm}(0) \vert k +3m \rangle$. In addition, by adding the first six equations of Eq.~(\ref{sistemaEc}), it can be verified that the derivative with respect to time of $\sum_{\alpha=\pm}\sum_{j=1}^{3}\lambda_{j,j}^{(\alpha)}(t)$ is zero, which is in accordance with the fact that $\sum_{\alpha=\pm}\sum_{j=1}^{3}\lambda_{j,j}^{(\alpha)}(t) = 1$.

In order to explore the possibility of steady-state solutions $\tilde{\lambda}_{j,k}^{(\pm)}$  of the system of differential equations in Eq.~(\ref{sistemaEc}), one just has to take all of the derivatives on the lefthand side of Eq.~(\ref{sistemaEc}) equal to zero and replace $\lambda_{j,k}^{(\pm)}(t)$ by $\tilde{\lambda}_{j,k}^{(\pm)}$. This leads to a homogeneous linear system of algebraic equations with a solution space of dimension one. The free parameter that appears can be computed by imposing the condition $\sum_{\alpha=\pm}\sum_{j=1}^{3}\tilde{\lambda}_{j,j}^{(\alpha)} =  1$. The analytical expression obtained is quite lengthy and, thus, is not included here. In addition, it can be verified that the real parts of the nonzero eigenvalues of the coefficient matrix associated with the system of equations in Eq.~(\ref{sistemaEc}) are negative. Consequently, regardless of the initial conditions, in the long-time limit all the solutions $\lambda_{j,k}^{(\pm)}(t)$ tend to the steady-state solution $\tilde{\lambda}_{j,k}^{(\pm)}$. According to Eq.~(\ref{v}), the long-time limit of the velocity, $v_{\infty}$, is also well defined and unique.

Substituting the steady-state solution $\tilde{\lambda}_{j,k}^{(\pm)}$ in Eq.~(\ref{v}), one obtains after a lengthy calculation that
\begin{equation}
\label{vAnalitica}
\frac{v_{\infty}}{\omega L} = \frac{3\Delta(1-\Delta^{2})\tilde{\gamma}^{2}\tilde{\Gamma}F}{G} , 
\end{equation}
where we have introduced the dimensionless parameters $\Delta = (\gamma_{+} - \gamma_{-})/\gamma$, $\tilde{\gamma} = \gamma/\omega$, and $\tilde{\Gamma} = \Gamma/\omega$, with $\gamma = \gamma_{+} + \gamma_{-}$, and the dimensionless quantities
\begin{eqnarray}
\label{Qnum}
F
&=&   -\tilde{\Gamma}^{3}(2\tilde{\gamma} + \tilde{\Gamma})\Big[ (5-\Delta^{2})\tilde{\gamma}^{2} + 10 \tilde{\gamma}\tilde{\Gamma} + 4\tilde{\Gamma}^{2} \Big] \cr
&& -4\tilde{\Gamma}\Big[ 8\Delta^{2}\tilde{\gamma}^{3} + 10(1+\Delta^{2})\tilde{\gamma}^{2}\tilde{\Gamma} + 16\tilde{\gamma}\tilde{\Gamma}^{2} + 7\tilde{\Gamma}^{3} \Big] \cr
&& +32\Big[ (1+\Delta^{2})\tilde{\gamma}^{2} + \tilde{\gamma} \tilde{\Gamma} - \tilde{\Gamma}^{2} \Big]  + 64 
\end{eqnarray}
and 
\begin{eqnarray}
\label{Qdenom}
G 
&=& \tilde{\Gamma}^{5} (\tilde{\gamma} + \tilde{\Gamma})^{2} (2\tilde{\gamma} + \tilde{\Gamma})\left[ 4\tilde{\Gamma}^{2} + 12\tilde{\gamma}\tilde{\Gamma} + (9-\Delta^{2})\tilde{\gamma}^{2} \right] \cr
&& + 8\tilde{\Gamma}^{3}(\tilde{\gamma} + \tilde{\Gamma})\Big[  (9 + 7\Delta^{2})\tilde{\gamma}^{4}  + 3(13 + 3 \Delta^{2}) \tilde{\gamma}^{3} \tilde{\Gamma} \cr
&& + 2(29 + 2\Delta^{2})\tilde{\gamma}^{2}\tilde{\Gamma}^{2} +36\tilde{\gamma} \tilde{\Gamma}^{3}  +8\tilde{\Gamma}^{4} \Big] \cr
&& + 4 \tilde{\Gamma} \Big[ 2(5 +22\Delta^{2} + 5\Delta^{4})\tilde{\gamma}^{5} \cr
&& + (127 + 122\Delta^{2} + 7\Delta^{4}) \tilde{\gamma}^{4} \tilde{\Gamma} + 4( 103 + 37 \Delta^{2})\tilde{\gamma}^{3} \tilde{\Gamma}^{2} \cr
&& + (583 + 73\Delta^{2})\tilde{\gamma}^{2} \tilde{\Gamma}^{3} + 384\tilde{\gamma}\tilde{\Gamma}^{4} + 96\tilde{\Gamma}^{5}  \Big]  \cr
&& + 32\Big[ 3(1 - \Delta^{4})\tilde{\gamma}^{4} + 2(13 + 7\Delta^{2})\tilde{\gamma}^{3}\tilde{\Gamma} \cr
&& + (71 + 17\Delta^{2})\tilde{\gamma}^{2}\tilde{\Gamma}^{2} + 80\tilde{\gamma}\tilde{\Gamma}^{3} +32 \tilde{\Gamma}^{4}  \Big] \cr
&& + 64\Big[ 3(1 - \Delta^{2})\tilde{\gamma}^{2} + 16\tilde{\gamma}\tilde{\Gamma} +16 \tilde{\Gamma}^{2} \Big] .
\end{eqnarray}
Notice that the denominator $G$ is always a positive quantity because $\vert \Delta \vert \leq 1$. 

In order to simplify the above complex expression, two asymptotic behaviors of interest can be considered. These asymptotic behaviors correspond to the cases where the \textit{coherent frequency} $\omega$ dominates over the \textit{incoherent rates} $\gamma$ and $\Gamma$ or viceversa. Formally, these correspond to taking either of the limits $\tilde{\gamma}, \tilde{\Gamma} \to 0$ or $\tilde{\gamma}, \tilde{\Gamma} \to \infty$, while keeping $g=\tilde{\Gamma}/ \tilde{\gamma} = \Gamma/\gamma$ fixed. In the first case (i.e., for $\gamma , \Gamma \ll \omega$), one obtains 
	\begin{equation}
		\label{lim1}
		\frac{v_{\infty}}{\omega L} \sim \frac{3 \Delta (1-\Delta^2)\tilde{\Gamma}}{(1+4g)(3+4g)-3\Delta^2} ,
	\end{equation}	
	while in the second (i.e., for $\gamma, \Gamma \gg \omega$), one finds  
	\begin{equation}
		\label{lim2}
		\frac{v_{\infty}}{\omega L} \sim \frac{3 \Delta (1-\Delta^2) \left[\Delta^2-5-2g(5+2g) \right]}{\tilde{\Gamma}(1+g)^2\left[(3+2g)^2-\Delta^2\right]} .
\end{equation}

	It is worth observing that $v_{\infty}$ can be alternatively expressed in terms of the mean residence times $T_{\pm}$ by setting $\gamma_{\pm} = T_{\pm}^{-1}$ in the definitions of $\Delta$ and $\tilde{\gamma}$. In particular, $\Delta$ can be interpreted as the difference of the mean residence times over their sum, $\Delta = (T_{-} - T_{+})/(T_{-} + T_{+})$, and  $\gamma $ as the sum of the reciprocals of the mean residence times, $\gamma = T_{+}^{-1} + T_{-}^{-1}$. In particular, $T_+=T_-$ implies that $\Delta=0$ and, consequently, that the mean average velocity cancels. In addition, if $T_+\gg T_-$ or $T_+\ll T_-$, then $\Delta \approx -1$ or $\Delta \approx +1$, respectively, and the mean average velocity also tends to cancel.

\section{Analysis of the long-time limit velocity}
\label{Spropiedades}

According to Eq.~(\ref{vAnalitica}), $v_{\infty}$ is an odd function of $\Delta$ and, consequently, vanishes when $\Delta = 0$. This corroborates the symmetry arguments presented in Sec.~\ref{SecV}, where it was shown that $v_{\infty}(\gamma_{+},\gamma_{-}) = - v_{\infty}(\gamma_{-},\gamma_{+})$ and, therefore, different transition rates are necessary to obtain directed motion. In addition, $v_{\infty} $ also vanishes when $\Delta = \pm 1$, which implies that both transition rates $\gamma_{\pm}$ must be nonzero to have directed motion. This is so because, in the absence of repetitive fluctuations between the two lattice states, the absorbing sites eventually trap the particle avoiding its motion.

In addition to the previously mentioned values of $\Delta$, directed motion can also be canceled for those values of $\Delta$, if any, for which $F=0$. The function $F$ depends on $\tilde{\gamma}$, $\tilde{\Gamma}$, and $\Delta^{2}$, i.e., $F = F(\tilde{\gamma}, \tilde{\Gamma}, \Delta^{2})$. 

\begin{figure}
	\centering
	\includegraphics[scale=0.45]{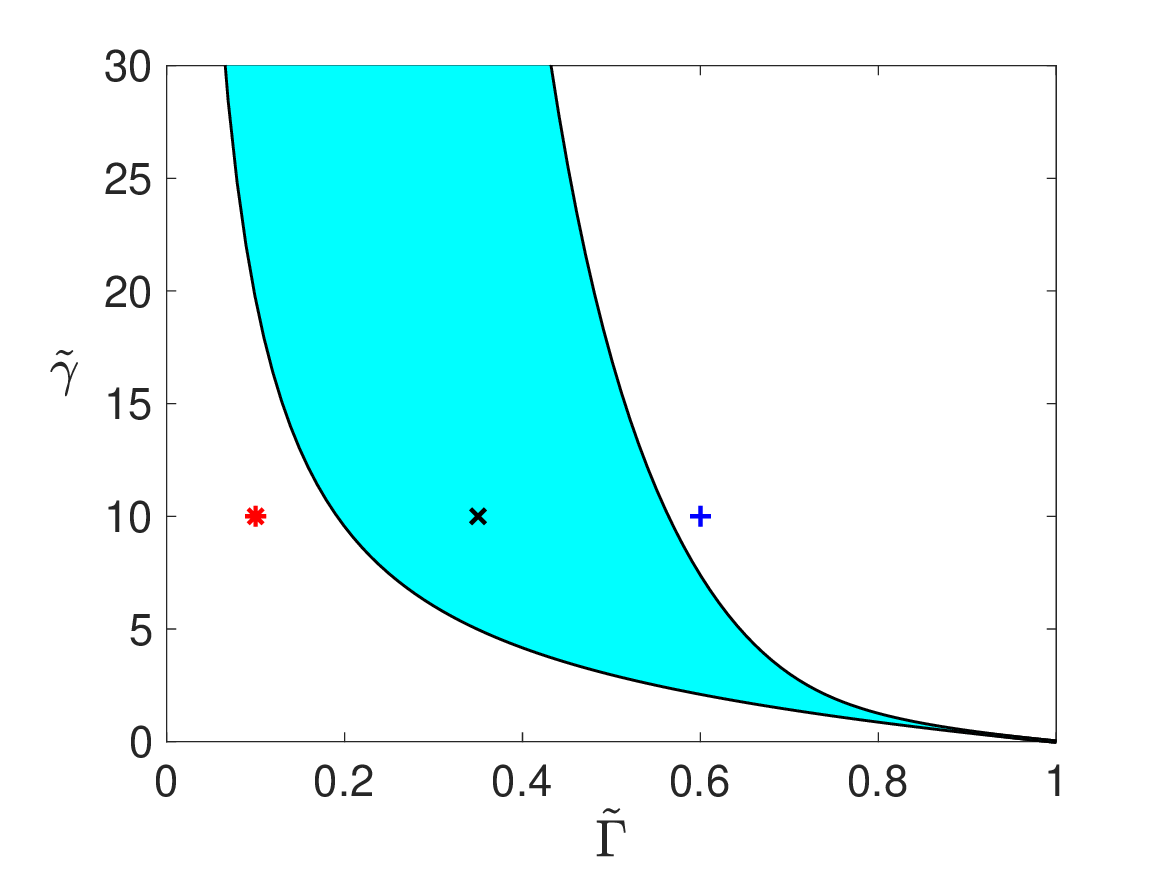}
	\caption{Classification of points $(\tilde{\Gamma},\tilde{\gamma})$ according to the number of directed-motion inversions exhibited by $v_{\infty}$ as a function of $\Delta$. The cyan-shaded area specifies the region where inversions of directed motion occur for three values of $\Delta$. Outside this region the only inversion occurs for $\Delta = 0$. The curves bounding the cyan-shaded area from above and from below arise from the conditions $F(\tilde{\gamma},\tilde{\Gamma}, 0) = 0$ and $F(\tilde{\gamma},\tilde{\Gamma}, 1) = 0$, respectively, with $F$ defined in Eq.~(\ref{Qnum}).  The red asterisk, the black cross, and the blue plus sign indicate the values of $(\tilde{\Gamma}, \tilde{\gamma})$ considered in Fig.~\ref{Figure3}. }     
	\label{Figure2}
\end{figure}

Since $F$ is a first-degree polynomial in $\Delta^{2}$, there is always a value $\Delta_{\mathrm{c}}^{2}$ that vanishes $F$. According to the definition of $\Delta$, only the values $\Delta_{\mathrm{c}}^{2}$ between $0$ and $1$ are acceptable. In Fig.~\ref{Figure2}, the cyan-shaded area depicts the region of points $(\tilde{\gamma}, \tilde{\Gamma})$ for which $\Delta_{\mathrm{c}}^{2} \in (0,1)$.  For each point in this region there exist three values of $\Delta$ where inversion of directed motion occurs, namely, $\Delta = 0$ and $\Delta = \pm \Delta_{\mathrm{c}}$. Outside of this region inversion of directed motion occurs only for $\Delta = 0$. The boundaries of this region are implicitly defined by the conditions $F(\tilde{\gamma},\tilde{\Gamma}, 1) = 0$ and $F(\tilde{\gamma},\tilde{\Gamma}, 0) = 0$, which correspond respectively to $\Delta_{\mathrm{c}}^{2} = 1$ and $\Delta_{\mathrm{c}}^{2} = 0$. These conditions are polynomial equations of degree three in $\tilde{\gamma}$ and can be solved analytically to obtain the dependence of $\tilde{\gamma}$ on $\tilde{\Gamma}$ for the curves that parametrize these boundaries. In particular, condition $F(\tilde{\gamma},\tilde{\Gamma}, 1) = 0$ yields $\tilde{\gamma}_{1}(\tilde{\Gamma}) = 2(1-\tilde{\Gamma}^{2})/\tilde{\Gamma}$, which is the curve that bounds the shaded area from below in Fig.~\ref{Figure2}. The curve $\tilde{\gamma}_{0}(\tilde{\Gamma})$ obtained from solving condition $F(\tilde{\gamma},\tilde{\Gamma}, 0) = 0$ bounds the shaded area from above. The analytical expression for $\tilde{\gamma}_{0}(\tilde{\Gamma})$ is very lengthy and, for that reason, is not included here. Since $\tilde{\gamma} \geq 0$, one conclusion that readily follows from the above expression for $\tilde{\gamma}_{1}(\tilde{\Gamma})$ is that inversion of directed motion for $\Delta \not= 0$ occurs only when $\tilde{\Gamma} < 1$.

\begin{figure}
   \centering
   \includegraphics[scale=0.45]{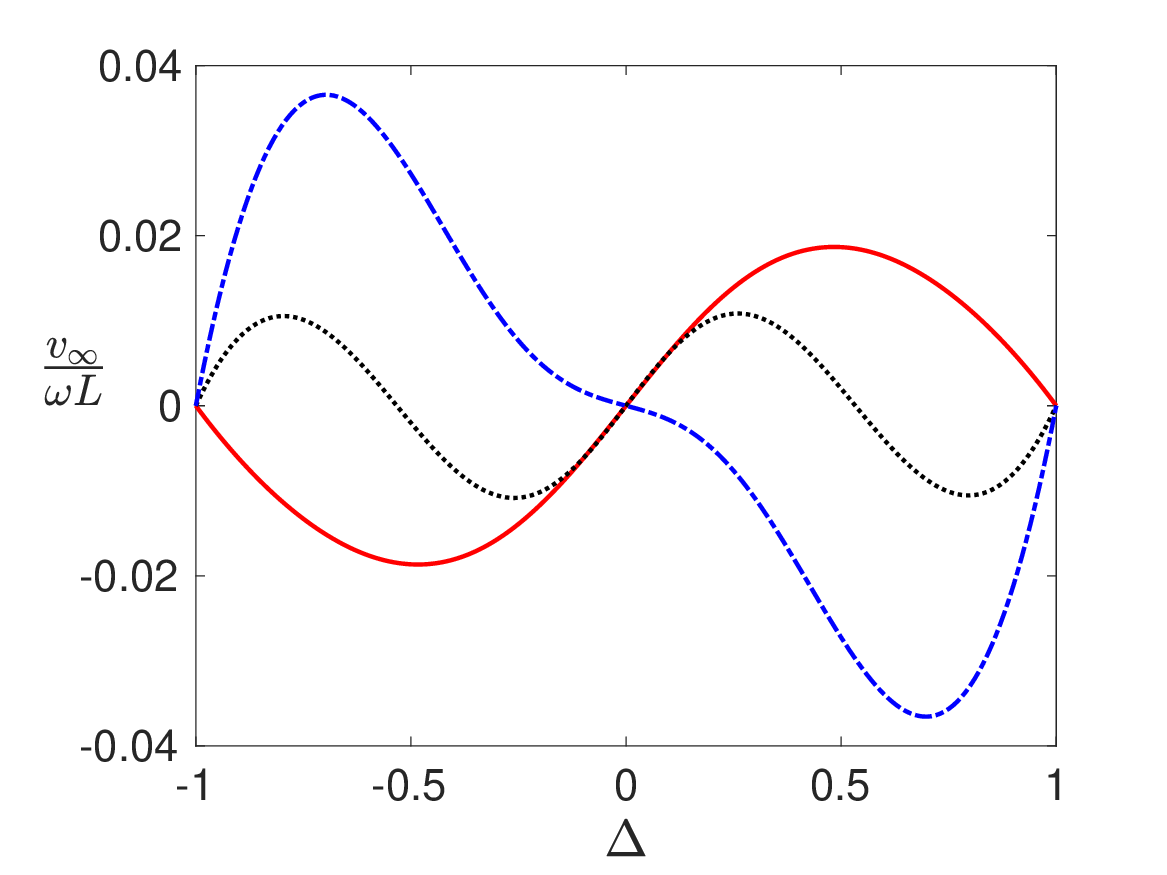}
   \caption{Dimensionless steady-state velocity $v_{\infty}/(\omega L)$ in Eq.~(\ref{vAnalitica}) as a function of $\Delta$ for $\tilde{\gamma} = 10$ and $\tilde{\Gamma} = 0.1$ (red solid line), $0.35$ (black dotted line), $0.6$ (blue dot-dashed line). The corresponding points $(\tilde{\gamma}, \tilde{\Gamma})$ are shown in Fig.~\ref{Figure2} as a red asterisk, a black cross, and a blue plus sign.}     
   \label{Figure3}
\end{figure}

In order to illustrate the inversion of directed motion, Fig.~\ref{Figure3} shows the dependence of the dimensionless long-time velocity $v_{\infty}/(\omega L)$ on the parameter $\Delta$. The values of $\tilde{\gamma}$ and $\tilde{\Gamma}$ have been chosen to lie in the three different regions shown in Fig.~\ref{Figure2}. These values are depicted in Fig.~\ref{Figure2} as a red asterisk, a black cross, and a blue plus sign. Notice that the black dotted curve exhibits inversion of directed motion at three values of $\Delta$, while the red solid and blue dot-dashed curves present inversion only at $\Delta = 0$. Moreover, observe that $v_{\infty}$ has opposite signs on the red solid and blue dot-dashed curves. Specifically, $\Delta v_{\infty} \geq 0$ if $(\tilde{\gamma}, \tilde{\Gamma})$ lies to the left of the cyan-shaded area in Fig.~\ref{Figure2}, whereas $\Delta v_{\infty} \leq 0$ if $(\tilde{\gamma}, \tilde{\Gamma})$ lies to the right of the aforementioned area. Notice that, in the asymptotic limits in Eqs.~(\ref{lim1}) and (\ref{lim2}), the multiple inversions of directed motion disappear. This is so because for these asymptotic expressions to be valid both $\tilde{\gamma}$ and $\tilde{\Gamma}$ have to be simultaneously much greater or much smaller than $1$. This condition is clearly not fulfilled in the cyan-shaded area depicted in Fig.~\ref{Figure2}.

\section{Conclusions}
\label{conclusiones}

In summary, this paper presented a simple quantum random-walk model on a one-dimensional periodic lattice that fluctuates between two distinct states.  The mathematical framework for this model is rooted in Lindblad rate equations, which encompass the transition rates between the lattice states. Despite the system's infinite-dimensional state space, the inherent periodic symmetry of the problem enables us to describe the time evolution of the particle velocity through a finite set of differential equations. These equations yield concise, analytical expressions for the long-time velocity behavior.

The simplicity of the derived analytical expression for the velocity in the long-time limit has enabled us to conduct a comprehensive investigation of the characteristics of directed motion. In particular, some of these properties emerge as direct consequences of the system's symmetries. The resultant velocity of the particle and the complexities observed, including multiple changes in motion direction, stem from the interplay between the quantum and semiclassical terms, $v_{\mathrm{q}}(t)$ and $v_{\mathrm{c}}(t)$,  composing the velocity in Eq.~(\ref{v}). These intricacies are notably absent in the fully classical model analyzed in Ref.~[10].
 
Finally, it is important to mention that, although the model presented in this work is quite simple, we are confident that some of the techniques presented here can be extended to more complex quantum models. Specifically, the use of the Lindblad rate equations proposed in this work opens up new perspectives for the study of other fluctuating quantum systems.

\acknowledgements

The authors acknowledge Grant No.~PID2022-136228NB-C22 funded by MICIU/AEI/10.13039/501100011033 and by ERDF, EU.  {The authors express their sincere gratitude to Eduardo G\'omez and Daniel Barredo for valuable discussions and guidance regarding a potential experimental realization of our model.

{

\appendix

\section{Derivation of the Lindblad rate equations~(\ref{evolucion})}
\label{ApendiceDerivacion}

Let us consider a quantum system evolving according to the Lindblad equation 
\begin{equation}
\label{EcAp1}
\frac{d}{dt}\rho (t) = \mathcal{L}_{\eta(t)} \rho(t) ,
\end{equation}
where the generator $\mathcal{L}_{\eta(t)}$ depends  on a dichotomic Markov process $\eta (t)$ that can take the values $\pm 1$. For simplicity, we write $\pm$ instead of $\pm 1$ when $\pm 1$ appears as a subscript or superscript, so that $\mathcal{L}_{\pm} \equiv \mathcal{L}_{\pm1}$. The process $\eta(t)$ is described by the master equation
\begin{equation}
\label{EcAp2}
\frac{\partial}{\partial t} p(\alpha , t \vert \alpha_{0}, t_{0}) = \gamma_{-\alpha}p(-\alpha , t \vert \alpha_{0}, t_{0}) - \gamma_{\alpha}p(\alpha, t \vert \alpha_{0}, t_{0}) ,
\end{equation}
where $p(\alpha , t \vert \alpha_{0}, t_{0})$ is the conditional probability that $\eta (t) = \alpha$ given that $\eta(t_{0}) = \alpha_{0}$, and $\gamma_{\alpha}$ is the transition rate from the value $\alpha$ to $-\alpha$. In the above expression, the parameters $\alpha$ and $\alpha_{0}$ can take the values $\pm 1$.

A realization of the process $\eta(t)$ in the time interval $[t_{0},t]$ can be characterized by giving its final value, more specifically, $\alpha = \lim_{t' \to t^{-}} \eta(t')$ and the time instants, if any, at which changes in value occur. These time instants are assumed to be listed in increasing order, i.e., $t_{0} < t_{1} < t_{2} < \cdots < t_{n} < t$, and arranged in a vector $\boldsymbol{\tau}_{n} = (t_{1},t_{2}, \dots, t_{n})$. For notational convenience, we define $\boldsymbol{\tau}_{0}$ as a vector with no components.

The probability of a realization with no changes in value is given by
\begin{equation}
\label{p0}
P_{0}^{[t_{0},t]}(\alpha,\boldsymbol{\tau}_{0}) = e^{-(t-t_{0})\gamma_{\alpha}}p(\alpha, t_{0}) ,
\end{equation}
where $p(\alpha, t_{0})$ is the probability that the process $\eta(t)$ takes the value $\alpha$ at time $t_{0}$. In this case, the density operator at time $t$ is
\begin{equation}
\label{rho0}
\rho_{0}(t;\alpha, \boldsymbol{\tau}_{0}) = e^{(t-t_{0})\mathcal{L}_{\alpha}} \rho(t_{0}) ,
\end{equation}
where $\rho(t_{0})$ is the density operator at the initial time $t_{0}$. The probability density of a realization ending in $\alpha$ with $n\geq 1$ changes in value is
\begin{equation}
\label{pn}
P_{n}^{[t_{0},t]}(\alpha,\boldsymbol{\tau}_{n}) = e^{-(t-t_{n})\gamma_{\alpha}} \prod_{j=1}^{n} \gamma_{\alpha_{j}} e^{-(t_{j}-t_{j-1})\gamma_{\alpha_{j}}} p(\alpha_{1}, t_{0})  ,
\end{equation}
where $\alpha_{j} = (-1)^{n+1-j}\alpha$ is the value of $\eta(t)$ just before the $j$th change. In this case, the density operator at time $t$ is given by
\begin{equation}
	\label{rhon}
	\rho_{n}(t;\alpha, \boldsymbol{\tau}_{n}) = e^{(t-t_{n})\mathcal{L}_{\alpha}} \prod_{j=1}^{n} e^{(t_{j}-t_{j-1})\mathcal{L}_{\alpha_{j}}} \rho(t_{0}) ,
\end{equation}
where the product of the superoperators must be taken in the order $\prod_{j=1}^{n} \mathcal{F}_{j} = \mathcal{F}_{n} \cdots \mathcal{F}_{1}$. Notice that the following recurrence relations hold for $n\geq 1$
\begin{equation}
	\label{recurrence1}
	P_{n}^{[t_{0},t]}(\alpha,\boldsymbol{\tau}_{n}) = \gamma_{-\alpha}e^{-(t-t_{n})\gamma_{\alpha}} P_{n-1}^{[t_{0},t_{n}]}(-\alpha, \boldsymbol{\tau}_{n-1})
\end{equation}	
and
\begin{equation}
\label{recurrence2}
	\rho_{n}(t;\alpha,\boldsymbol{\tau}_{n})  = e^{(t-t_{n})\mathcal{L}_{\alpha}}\rho_{n-1}(t_{n};-\alpha,\boldsymbol{\tau}_{n-1}) .
\end{equation}

The density operator at time $t$, $\rho(t)$, is obtained after an average over all realizations, i.e.,
\begin{equation}
\label{rhoF}
\rho(t) = \sum_{\alpha = \pm 1}  \rho_{\alpha}(t) , 
\end{equation}
where
\begin{eqnarray}
\label{rhoalfa}
\rho_{\alpha} (t) 
&=& \rho_{0}(t;\alpha,\boldsymbol{\tau}_{0})P_{0}^{[t_{0},t]}(\alpha,\boldsymbol{\tau}_{0}) \cr
&& + \sum_{n=1}^{\infty} \int_{t_{0}}^{t} d\boldsymbol{\tau}_{n} \rho_{n}(t;\alpha,\boldsymbol{\tau}_{n}) P_{n}^{[t_{0},t]}(\alpha,\boldsymbol{\tau}_{n}) , \quad\quad
\end{eqnarray}
and we have introduced the notation $\int_{t_{0}}^{t} d\boldsymbol{\tau}_{n} = \int_{t_{0}}^{t} dt_{n} \int_{t_{0}}^{t_{n}} dt_{n-1} \dots \int_{t_{0}}^{t_{2}} dt_{1}$.

Taking into account the recurrence relations in Eqs.~(\ref{recurrence1}) and (\ref{recurrence2}),  it can be verified that Eq.~(\ref{rhoalfa}) can be expressed as
\begin{eqnarray}
\label{rhoalfaB}
\rho_{\alpha}(t) 
&=& e^{-(t-t_{0})\gamma_{\alpha}} p(\alpha, t_{0}) e^{(t-t_{0})\mathcal{L}_{\alpha}} \rho(t_{0}) \cr
&& + \int_{t_{0}}^{t}dt' \gamma_{-\alpha}e^{-(t-t')\gamma_{\alpha}} e^{(t-t')\mathcal{L}_{\alpha}}\rho_{-\alpha}(t') . \quad\quad
\end{eqnarray}
Taking the derivative of Eq.~(\ref{rhoalfaB}) with respect to $t$, the evolution equations~(\ref{evolucion}) are finally obtained. In addition, from the integral equation~(\ref{rhoalfaB}), it readily follows that the initial conditions for these equations are $\rho_{\alpha}(t_{0}) = p(\alpha,t_{0})\rho(t_{0})$.

\section{Derivation of Eq.~(\ref{v})}
\label{ApendiceV}

Every integer $n$ can be expressed in the form $n = j + 3k$ with $j \in \{ 1, 2, 3 \}$ and $k \in \mathbb{Z}$. Therefore, using the translation superoperator $\mathcal{T}_{3k}$, one can write
\begin{eqnarray}
\label{EcApB1}
\mathrm{Tr}\left[ X \rho_{\pm}(t) \right]  
&=& L \sum_{n\in\mathbb{Z}} n \langle n \vert \rho_{\pm}(t) \vert n \rangle  \cr
&=& L \sum_{j = 1}^{3} \sum_{k\in\mathbb{Z}} (j + 3k) \langle j + 3k \vert \rho_{\pm}(t) \vert j + 3k \rangle  \cr
&=& L \sum_{j = 1}^{3} \sum_{k\in\mathbb{Z}} (j - 3k) \langle j \vert \mathcal{T}_{3k} \rho_{\pm}(t) \vert j \rangle .
\end{eqnarray}
Taking the derivative of the above expression with respect to time and using the evolution equations for $\rho_{\pm}(t)$ in Eq.~(\ref{evolucion}), as well as the definitions of the expected value of the position and the velocity in Eqs.~(\ref{X}) and (\ref{vd}), one obtains that
\begin{equation}
\label{EcApB2}
v(t) = L\sum_{\alpha = \pm}\sum_{j=1}^{3} \sum_{k\in\mathbb{Z}} (j-3k) \langle j \vert \mathcal{T}_{3k} \mathcal{L}_{\alpha}\rho_{\alpha}(t) \vert j \rangle .
\end{equation}
From the invariance of $\mathcal{L}_{\pm}$ with respect to translations by $3k$ units in Eq.~(\ref{simetria1}) and the definition of the operators $\Lambda_{\pm}(t)$ in Eq.~(\ref{19}), it then follows that 
\begin{equation}
\label{EcApB3}
\frac{v(t)}{L} = \sum_{\alpha = \pm}\sum_{j=1}^{3} \sum_{k\in\mathbb{Z}} (j-3k) \langle j \vert  \mathcal{L}_{\alpha}\mathcal{T}_{3k}\rho_{\alpha}(t) \vert j \rangle = A + B ,
\end{equation}
where
\begin{equation}
A = \sum_{\alpha = \pm} \sum_{j=1}^{3} j \langle j \vert \mathcal{L}_{\alpha}\Lambda_{\alpha}(t) \vert j \rangle 
\end{equation}
and
\begin{equation}
B = -3\sum_{\alpha = \pm}\sum_{k\in\mathbb{Z}} k \sum_{j=1}^{3} \langle j \vert \mathcal{L}_{\alpha}\mathcal{T}_{3k} \rho_{\alpha}(t) \vert j \rangle .
\end{equation}
To evaluate $A$ and $B$ one has to apply the definitions of the superoperators $\mathcal{L}_{\pm}$ in Eqs.~(\ref{LkLkp})-(\ref{disipadorm}) and the periodicity of the matrix elements of $\Lambda_{\pm}(t)$ in Eq.~(\ref{periodicidad}). It can then be verified from straightforward calculations that 
\begin{eqnarray}
\label{EcApB4}
A &=& \Gamma\left[ 2\lambda_{1,1}^{(+)}(t) + \lambda_{2,2}^{(+)}(t) - 2\lambda_{3,3}^{(-)}(t) - \lambda_{2,2}^{(-)}(t) \right]  \cr
&& - 2\omega\mathrm{Im}\left[ \lambda_{2,1}^{(+)}(t) + \lambda_{3,2}^{(-)}(t) \right] 
\end{eqnarray}
and
\begin{eqnarray}
\label{EcApB5}
B
&=& -3\Gamma \sum_{k\in\mathbb{Z}} k \Big[ \langle 4 \vert \mathcal{T}_{3k}\rho_{+}(t) \vert 4 \rangle - \langle 1 \vert \mathcal{T}_{3k}\rho_{+}(t) \vert 1 \rangle \cr
 && \quad\quad\quad + \langle 0 \vert \mathcal{T}_{3k}\rho_{-}(t) \vert 0 \rangle - \langle 3 \vert \mathcal{T}_{3k}\rho_{-}(t) \vert 3 \rangle\Big] .
\end{eqnarray}
Taking into account the definition of the translation superoperators, it is clear that $\langle 4 \vert \mathcal{T}_{3k}\rho_{+}(t) \vert 4 \rangle = \langle 1 \vert \mathcal{T}_{3(k-1)}\rho_{+}(t) \vert 1 \rangle$ and $\langle 0 \vert \mathcal{T}_{3k}\rho_{-}(t) \vert 0 \rangle = \langle 3 \vert \mathcal{T}_{3(k+1)}\rho_{+}(t) \vert 3 \rangle$. A change of summation index in Eq.~(\ref{EcApB5}) leads to the result $B = -3\Gamma[\lambda_{1,1}^{(+)}(t) - \lambda_{3,3}^{(-)}(t)]$. By adding the above results for $A$ and $B$,  the expression for the velocity in Eq.~(\ref{v}) is finally obtained.

\providecommand{\noopsort}[1]{}\providecommand{\singleletter}[1]{#1}%
%


\begin{thebibliography}{40}%
	\makeatletter
	\providecommand \@ifxundefined [1]{%
		\@ifx{#1\undefined}
	}%
	\providecommand \@ifnum [1]{%
		\ifnum #1\expandafter \@firstoftwo
		\else \expandafter \@secondoftwo
		\fi
	}%
	\providecommand \@ifx [1]{%
		\ifx #1\expandafter \@firstoftwo
		\else \expandafter \@secondoftwo
		\fi
	}%
	\providecommand \natexlab [1]{#1}%
	\providecommand \enquote  [1]{``#1''}%
	\providecommand \bibnamefont  [1]{#1}%
	\providecommand \bibfnamefont [1]{#1}%
	\providecommand \citenamefont [1]{#1}%
	\providecommand \href@noop [0]{\@secondoftwo}%
	\providecommand \href [0]{\begingroup \@sanitize@url \@href}%
	\providecommand \@href[1]{\@@startlink{#1}\@@href}%
	\providecommand \@@href[1]{\endgroup#1\@@endlink}%
	\providecommand \@sanitize@url [0]{\catcode `\\12\catcode `\$12\catcode
		`\&12\catcode `\#12\catcode `\^12\catcode `\_12\catcode `\%12\relax}%
	\providecommand \@@startlink[1]{}%
	\providecommand \@@endlink[0]{}%
	\providecommand \url  [0]{\begingroup\@sanitize@url \@url }%
	\providecommand \@url [1]{\endgroup\@href {#1}{\urlprefix }}%
	\providecommand \urlprefix  [0]{URL }%
	\providecommand \Eprint [0]{\href }%
	\providecommand \doibase [0]{https://doi.org/}%
	\providecommand \selectlanguage [0]{\@gobble}%
	\providecommand \bibinfo  [0]{\@secondoftwo}%
	\providecommand \bibfield  [0]{\@secondoftwo}%
	\providecommand \translation [1]{[#1]}%
	\providecommand \BibitemOpen [0]{}%
	\providecommand \bibitemStop [0]{}%
	\providecommand \bibitemNoStop [0]{.\EOS\space}%
	\providecommand \EOS [0]{\spacefactor3000\relax}%
	\providecommand \BibitemShut  [1]{\csname bibitem#1\endcsname}%
	\let\auto@bib@innerbib\@empty
	\bibitem [{\citenamefont {Reimann}(2002)}]{ReimannR}%
	\BibitemOpen
	\bibfield  {author} {\bibinfo {author} {\bibfnamefont {P.}~\bibnamefont
			{Reimann}},\ }\bibfield  {title} {\bibinfo {title} {Brownian motors: noisy
			transport far from equilibrium},\ }\href
	{https://doi.org/https://doi.org/10.1016/S0370-1573(01)00081-3} {\bibfield
		{journal} {\bibinfo  {journal} {Phys. Rep.}\ }\textbf {\bibinfo {volume}
			{361}},\ \bibinfo {pages} {57} (\bibinfo {year} {2002})}\BibitemShut
	{NoStop}%
	\bibitem [{\citenamefont {Ajdari}\ and\ \citenamefont
		{Prost}(1997)}]{MolecularMotors}%
	\BibitemOpen
	\bibfield  {author} {\bibinfo {author} {\bibfnamefont {F.~J.~A.}\
			\bibnamefont {Ajdari}}\ and\ \bibinfo {author} {\bibfnamefont
			{J.}~\bibnamefont {Prost}},\ }\bibfield  {title} {\bibinfo {title} {Modeling
			molecular motors},\ }\href
	{https://doi.org/https://doi.org/10.1103/RevModPhys.69.1269} {\bibfield
		{journal} {\bibinfo  {journal} {Rev. Mod. Phys.}\ }\textbf {\bibinfo {volume}
			{69}},\ \bibinfo {pages} {4} (\bibinfo {year} {1997})}\BibitemShut {NoStop}%
	\bibitem [{\citenamefont {H\"anggi}\ and\ \citenamefont
		{Marchesoni}(2009)}]{ArtificialBrownianMotors}%
	\BibitemOpen
	\bibfield  {author} {\bibinfo {author} {\bibfnamefont {P.}~\bibnamefont
			{H\"anggi}}\ and\ \bibinfo {author} {\bibfnamefont {F.}~\bibnamefont
			{Marchesoni}},\ }\bibfield  {title} {\bibinfo {title} {Artificial brownian
			motors: Controlling transport on the nanoscale},\ }\href
	{https://doi.org/10.1103/RevModPhys.81.387} {\bibfield  {journal} {\bibinfo
			{journal} {Rev. Mod. Phys.}\ }\textbf {\bibinfo {volume} {81}},\ \bibinfo
		{pages} {387} (\bibinfo {year} {2009})}\BibitemShut {NoStop}%
	\bibitem [{\citenamefont {Ghosh}\ and\ \citenamefont {Ray}(2007)}]{India}%
	\BibitemOpen
	\bibfield  {author} {\bibinfo {author} {\bibfnamefont {P.~K.}\ \bibnamefont
			{Ghosh}}\ and\ \bibinfo {author} {\bibfnamefont {D.~S.}\ \bibnamefont
			{Ray}},\ }\bibfield  {title} {\bibinfo {title} {Quantum ratchet motion},\
	}\href@noop {} {\bibfield  {journal} {\bibinfo  {journal} {J. Indian Inst.
				Sci.}\ }\textbf {\bibinfo {volume} {87}},\ \bibinfo {pages} {401} (\bibinfo
		{year} {2007})}\BibitemShut {NoStop}%
	\bibitem [{\citenamefont {Vaquero-Stainer}\ \emph {et~al.}(2018)\citenamefont
		{Vaquero-Stainer}, \citenamefont {Yoshida}, \citenamefont {Hylton},
		\citenamefont {Pusch}, \citenamefont {Curtin}, \citenamefont {Frogley},
		\citenamefont {Wilson}, \citenamefont {Clarke}, \citenamefont {Kennedy},
		\citenamefont {Ekins-Daukes}, \citenamefont {Hess},\ and\ \citenamefont
		{Phillips}}]{CommPhys}%
	\BibitemOpen
	\bibfield  {author} {\bibinfo {author} {\bibfnamefont {A.}~\bibnamefont
			{Vaquero-Stainer}}, \bibinfo {author} {\bibfnamefont {M.}~\bibnamefont
			{Yoshida}}, \bibinfo {author} {\bibfnamefont {N.~P.}\ \bibnamefont {Hylton}},
		\bibinfo {author} {\bibfnamefont {A.}~\bibnamefont {Pusch}}, \bibinfo
		{author} {\bibfnamefont {O.}~\bibnamefont {Curtin}}, \bibinfo {author}
		{\bibfnamefont {M.}~\bibnamefont {Frogley}}, \bibinfo {author} {\bibfnamefont
			{T.}~\bibnamefont {Wilson}}, \bibinfo {author} {\bibfnamefont
			{E.}~\bibnamefont {Clarke}}, \bibinfo {author} {\bibfnamefont
			{K.}~\bibnamefont {Kennedy}}, \bibinfo {author} {\bibfnamefont {N.~J.}\
			\bibnamefont {Ekins-Daukes}}, \bibinfo {author} {\bibfnamefont
			{O.}~\bibnamefont {Hess}},\ and\ \bibinfo {author} {\bibfnamefont {C.~C.}\
			\bibnamefont {Phillips}},\ }\bibfield  {title} {\bibinfo {title}
		{Semiconductor nanostructure quantum ratchet for high efficiency solar
			cells},\ }\href@noop {} {\bibfield  {journal} {\bibinfo  {journal} {Comm.
				Phys.}\ }\textbf {\bibinfo {volume} {1}} (\bibinfo {year}
		{2018})}\BibitemShut {NoStop}%
	\bibitem [{\citenamefont {Marchesoni}(1996)}]{Marchesoni1996}%
	\BibitemOpen
	\bibfield  {author} {\bibinfo {author} {\bibfnamefont {F.}~\bibnamefont
			{Marchesoni}},\ }\bibfield  {title} {\bibinfo {title} {Thermal ratchets in
			$1+1$ dimensions},\ }\href {https://doi.org/10.1103/physrevlett.77.2364}
	{\bibfield  {journal} {\bibinfo  {journal} {Phys. Rev. Lett.}\ }\textbf
		{\bibinfo {volume} {77}},\ \bibinfo {pages} {2364} (\bibinfo {year}
		{1996})}\BibitemShut {NoStop}%
	\bibitem [{\citenamefont {Salerno}\ and\ \citenamefont
		{Quintero}(2002)}]{Salerno2002}%
	\BibitemOpen
	\bibfield  {author} {\bibinfo {author} {\bibfnamefont {M.}~\bibnamefont
			{Salerno}}\ and\ \bibinfo {author} {\bibfnamefont {N.~R.}\ \bibnamefont
			{Quintero}},\ }\bibfield  {title} {\bibinfo {title} {Soliton ratchets},\
	}\href@noop {} {\bibfield  {journal} {\bibinfo  {journal} {Phys. Rev. E}\
		}\textbf {\bibinfo {volume} {65}},\ \bibinfo {pages} {025602(R)} (\bibinfo
		{year} {2002})}\BibitemShut {NoStop}%
	\bibitem [{\citenamefont {Salerno}\ and\ \citenamefont
		{Zolotaryuk}(2002)}]{Salerno2002b}%
	\BibitemOpen
	\bibfield  {author} {\bibinfo {author} {\bibfnamefont {M.}~\bibnamefont
			{Salerno}}\ and\ \bibinfo {author} {\bibfnamefont {Y.}~\bibnamefont
			{Zolotaryuk}},\ }\bibfield  {title} {\bibinfo {title} {Soliton ratchetlike
			dynamics by ac forces with harmonic mixing},\ }\href@noop {} {\bibfield
		{journal} {\bibinfo  {journal} {Phys. Rev. E}\ }\textbf {\bibinfo {volume}
			{65}},\ \bibinfo {pages} {056603} (\bibinfo {year} {2002})}\BibitemShut
	{NoStop}%
	\bibitem [{\citenamefont {Morales-Molina}\ \emph {et~al.}(2003)\citenamefont
		{Morales-Molina}, \citenamefont {Mertens},\ and\ \citenamefont
		{S\'{a}nchez}}]{MoralesMolina2003}%
	\BibitemOpen
	\bibfield  {author} {\bibinfo {author} {\bibfnamefont {L.}~\bibnamefont
			{Morales-Molina}}, \bibinfo {author} {\bibfnamefont {F.~G.}\ \bibnamefont
			{Mertens}},\ and\ \bibinfo {author} {\bibfnamefont {A.}~\bibnamefont
			{S\'{a}nchez}},\ }\bibfield  {title} {\bibinfo {title} {Soliton ratchets out
			of point-like inhomogeneities},\ }\href
	{https://doi.org/10.1140/epjb/e2004-00031-3} {\bibfield  {journal} {\bibinfo
			{journal} {Eur. Phys. J. B}\ }\textbf {\bibinfo {volume} {37}},\ \bibinfo
		{pages} {79} (\bibinfo {year} {2003})}\BibitemShut {NoStop}%
	\bibitem [{\citenamefont {Morales-Molina}\ \emph {et~al.}(2005)\citenamefont
		{Morales-Molina}, \citenamefont {Mertens},\ and\ \citenamefont
		{S\'anchez}}]{MoralesMolina2005}%
	\BibitemOpen
	\bibfield  {author} {\bibinfo {author} {\bibfnamefont {L.}~\bibnamefont
			{Morales-Molina}}, \bibinfo {author} {\bibfnamefont {F.~G.}\ \bibnamefont
			{Mertens}},\ and\ \bibinfo {author} {\bibfnamefont {A.}~\bibnamefont
			{S\'anchez}},\ }\bibfield  {title} {\bibinfo {title} {Ratchet behavior in
			nonlinear {K}lein-{G}ordon systems with pointlike inhomogeneities},\
	}\href@noop {} {\bibfield  {journal} {\bibinfo  {journal} {Phys. Rev. E}\
		}\textbf {\bibinfo {volume} {72}},\ \bibinfo {pages} {016612} (\bibinfo
		{year} {2005})}\BibitemShut {NoStop}%
	\bibitem [{\citenamefont {S\'anchez-Rey}\ \emph {et~al.}(2016)\citenamefont
		{S\'anchez-Rey}, \citenamefont {Casado-Pascual},\ and\ \citenamefont
		{Quintero}}]{SnchezRey2016}%
	\BibitemOpen
	\bibfield  {author} {\bibinfo {author} {\bibfnamefont {B.}~\bibnamefont
			{S\'anchez-Rey}}, \bibinfo {author} {\bibfnamefont {J.}~\bibnamefont
			{Casado-Pascual}},\ and\ \bibinfo {author} {\bibfnamefont {N.~R.}\
			\bibnamefont {Quintero}},\ }\bibfield  {title} {\bibinfo {title} {Kink
			ratchet induced by a time-dependent symmetric field potential},\ }\href@noop
	{} {\bibfield  {journal} {\bibinfo  {journal} {Phys. Rev. E}\ }\textbf
		{\bibinfo {volume} {94}},\ \bibinfo {pages} {012221} (\bibinfo {year}
		{2016})}\BibitemShut {NoStop}%
	\bibitem [{\citenamefont {Casado-Pascual}\ \emph {et~al.}(2019)\citenamefont
		{Casado-Pascual}, \citenamefont {S\'{a}nchez-Rey},\ and\ \citenamefont
		{Quintero}}]{CasadoPascual2019}%
	\BibitemOpen
	\bibfield  {author} {\bibinfo {author} {\bibfnamefont {J.}~\bibnamefont
			{Casado-Pascual}}, \bibinfo {author} {\bibfnamefont {B.}~\bibnamefont
			{S\'{a}nchez-Rey}},\ and\ \bibinfo {author} {\bibfnamefont {N.~R.}\
			\bibnamefont {Quintero}},\ }\bibfield  {title} {\bibinfo {title} {Soliton
			ratchet induced by random transitions among symmetric sine-{G}ordon
			potentials},\ }\href@noop {} {\bibfield  {journal} {\bibinfo  {journal}
			{Chaos}\ }\textbf {\bibinfo {volume} {29}},\ \bibinfo {pages} {053119} (\bibinfo {year}
		{2019})}\BibitemShut {NoStop}%
	\bibitem [{\citenamefont {Reimann}(2001)}]{Reimann2001}%
	\BibitemOpen
	\bibfield  {author} {\bibinfo {author} {\bibfnamefont {P.}~\bibnamefont
			{Reimann}},\ }\bibfield  {title} {\bibinfo {title} {Supersymmetric
			ratchets},\ }\href {https://doi.org/10.1103/physrevlett.86.4992} {\bibfield
		{journal} {\bibinfo  {journal} {Phys. Rev. Lett.}\ }\textbf {\bibinfo
			{volume} {86}},\ \bibinfo {pages} {4992} (\bibinfo {year}
		{2001})}\BibitemShut {NoStop}%
	\bibitem [{\citenamefont {Doering}\ \emph {et~al.}(1994)\citenamefont
		{Doering}, \citenamefont {Horsthemke},\ and\ \citenamefont
		{Riordan}}]{Doering1994}%
	\BibitemOpen
	\bibfield  {author} {\bibinfo {author} {\bibfnamefont {C.~R.}\ \bibnamefont
			{Doering}}, \bibinfo {author} {\bibfnamefont {W.}~\bibnamefont
			{Horsthemke}},\ and\ \bibinfo {author} {\bibfnamefont {J.}~\bibnamefont
			{Riordan}},\ }\bibfield  {title} {\bibinfo {title} {Nonequilibrium
			fluctuation-induced transport},\ }\href
	{https://doi.org/10.1103/physrevlett.72.2984} {\bibfield  {journal} {\bibinfo
			{journal} {Phys. Rev. Lett.}\ }\textbf {\bibinfo {volume} {72}},\ \bibinfo
		{pages} {2984} (\bibinfo {year} {1994})}\BibitemShut {NoStop}%
	\bibitem [{\citenamefont {Elston}\ and\ \citenamefont
		{Doering}(1996)}]{Elston1996}%
	\BibitemOpen
	\bibfield  {author} {\bibinfo {author} {\bibfnamefont {T.~C.}\ \bibnamefont
			{Elston}}\ and\ \bibinfo {author} {\bibfnamefont {C.~R.}\ \bibnamefont
			{Doering}},\ }\bibfield  {title} {\bibinfo {title} {Numerical and analytical
			studies of nonequilibrium fluctuation-induced transport processes},\ }\href
	{https://doi.org/10.1007/bf02183737} {\bibfield  {journal} {\bibinfo
			{journal} {J. Stat. Phys.}\ }\textbf {\bibinfo {volume} {83}},\ \bibinfo
		{pages} {359} (\bibinfo {year} {1996})}\BibitemShut {NoStop}%
	\bibitem [{\citenamefont {Casado-Pascual}(2006)}]{Casado2006}%
	\BibitemOpen
	\bibfield  {author} {\bibinfo {author} {\bibfnamefont {J.}~\bibnamefont
			{Casado-Pascual}},\ }\bibfield  {title} {\bibinfo {title} {Flux reversal in a
			simple random-walk model on a fluctuating symmetric lattice},\ }\href
	{https://doi.org/https://doi.org/10.1103/PhysRevLett.131.133401} {\bibfield
		{journal} {\bibinfo  {journal} {Phys. Rev. E}\ }\textbf {\bibinfo {volume}
			{74}},\ \bibinfo {pages} {021112} (\bibinfo {year} {2006})}\BibitemShut
	{NoStop}%
	\bibitem [{\citenamefont {Casado-Pascual}(2018)}]{Casado2018}%
	\BibitemOpen
	\bibfield  {author} {\bibinfo {author} {\bibfnamefont {J.}~\bibnamefont
			{Casado-Pascual}},\ }\bibfield  {title} {\bibinfo {title} {Directed motion of
			spheres induced by unbiased driving forces in viscous fluids beyond the
			{S}tokes’ law regime},\ }\href {https://doi.org/DOI:
		10.1103/PhysRevE.97.032219} {\bibfield  {journal} {\bibinfo  {journal} {Phys.
				Rev. E}\ }\textbf {\bibinfo {volume} {97}},\ \bibinfo {pages} {032219}
		(\bibinfo {year} {2018})}\BibitemShut {NoStop}%
	\bibitem [{\citenamefont {Reimann}\ \emph {et~al.}(1997)\citenamefont
		{Reimann}, \citenamefont {Grifoni},\ and\ \citenamefont {H\"anggi}}]{QR1997}%
	\BibitemOpen
	\bibfield  {author} {\bibinfo {author} {\bibfnamefont {P.}~\bibnamefont
			{Reimann}}, \bibinfo {author} {\bibfnamefont {M.}~\bibnamefont {Grifoni}},\
		and\ \bibinfo {author} {\bibfnamefont {P.}~\bibnamefont {H\"anggi}},\
	}\bibfield  {title} {\bibinfo {title} {Quantum ratchets},\ }\href@noop {}
	{\bibfield  {journal} {\bibinfo  {journal} {Phys. Rev. Lett.}\ }\textbf
		{\bibinfo {volume} {79}},\ \bibinfo {pages} {10} (\bibinfo {year}
		{1997})}\BibitemShut {NoStop}%
	\bibitem [{\citenamefont {Yukawa}\ \emph {et~al.}(1997)\citenamefont {Yukawa},
		\citenamefont {Kikuchi}, \citenamefont {Tatara},\ and\ \citenamefont
		{Matsukawa}}]{Yukawa1997}%
	\BibitemOpen
	\bibfield  {author} {\bibinfo {author} {\bibfnamefont {S.}~\bibnamefont
			{Yukawa}}, \bibinfo {author} {\bibfnamefont {M.}~\bibnamefont {Kikuchi}},
		\bibinfo {author} {\bibfnamefont {G.}~\bibnamefont {Tatara}},\ and\ \bibinfo
		{author} {\bibfnamefont {H.}~\bibnamefont {Matsukawa}},\ }\bibfield  {title}
	{\bibinfo {title} {Quantum ratchets},\ }\href
	{https://doi.org/10.1143/jpsj.66.2953} {\bibfield  {journal} {\bibinfo
			{journal} {J. Phys. Soc. Jpn.}\ }\textbf {\bibinfo {volume} {66}},\ \bibinfo
		{pages} {2953} (\bibinfo {year} {1997})}\BibitemShut {NoStop}%
	\bibitem [{\citenamefont {Ang}\ \emph {et~al.}(2015)\citenamefont {Ang},
		\citenamefont {Ma},\ and\ \citenamefont {Zhang}}]{SciRep2014}%
	\BibitemOpen
	\bibfield  {author} {\bibinfo {author} {\bibfnamefont {Y.~S.}\ \bibnamefont
			{Ang}}, \bibinfo {author} {\bibfnamefont {Z.}~\bibnamefont {Ma}},\ and\
		\bibinfo {author} {\bibfnamefont {C.}~\bibnamefont {Zhang}},\ }\bibfield
	{title} {\bibinfo {title} {Quantum ratchet in two-dimensional semiconductors
			with {R}ashba spin-orbit interaction},\ }\href {https://doi.org/DOI:
		10.1038/srep07872} {\bibfield  {journal} {\bibinfo  {journal} {Sci. Rep.}\
		}\textbf {\bibinfo {volume} {5}},\ \bibinfo {pages} {7872} (\bibinfo {year}
		{2015})}\BibitemShut {NoStop}%
	\bibitem [{\citenamefont {Zhan}\ \emph {et~al.}(2011)\citenamefont {Zhan},
		\citenamefont {Denisov}, \citenamefont {Ponomarev},\ and\ \citenamefont
		{H\"{a}nggi}}]{PRA842011}%
	\BibitemOpen
	\bibfield  {author} {\bibinfo {author} {\bibfnamefont {F.}~\bibnamefont
			{Zhan}}, \bibinfo {author} {\bibfnamefont {S.}~\bibnamefont {Denisov}},
		\bibinfo {author} {\bibfnamefont {A.~V.}\ \bibnamefont {Ponomarev}},\ and\
		\bibinfo {author} {\bibfnamefont {P.}~\bibnamefont {H\"{a}nggi}},\ }\bibfield
	{title} {\bibinfo {title} {Quantum ratchet transport with minimal dispersion
			rate},\ }\href {https://doi.org/DOI: 10.1103/PhysRevA.84.043617} {\bibfield
		{journal} {\bibinfo  {journal} {Phys. Rev. A}\ }\textbf {\bibinfo {volume}
			{84}},\ \bibinfo {pages} {043617} (\bibinfo {year} {2011})}\BibitemShut
	{NoStop}%
	\bibitem [{\citenamefont {Park}\ \emph {et~al.}(2019)\citenamefont {Park},
		\citenamefont {Ishizuka},\ and\ \citenamefont {Nagaosa}}]{PRB100}%
	\BibitemOpen
	\bibfield  {author} {\bibinfo {author} {\bibfnamefont {T.}~\bibnamefont
			{Park}}, \bibinfo {author} {\bibfnamefont {H.}~\bibnamefont {Ishizuka}},\
		and\ \bibinfo {author} {\bibfnamefont {N.}~\bibnamefont {Nagaosa}},\
	}\bibfield  {title} {\bibinfo {title} {Nonreciprocal transport of a
			super-{O}hmic quantum ratchet},\ }\href
	{https://doi.org/DOI:10.1103/PhysRevB.100.224301} {\bibfield  {journal}
		{\bibinfo  {journal} {Phys. Rev. B}\ }\textbf {\bibinfo {volume} {100}},\
		\bibinfo {pages} {224301} (\bibinfo {year} {2019})}\BibitemShut {NoStop}%
			\bibitem [{\citenamefont {Ungar}\ \emph {et~al.}(2019)\citenamefont {Ungar},
		\citenamefont {Cygorek},\ and\ \citenamefont {Axt}}]{PRB2019phonon}%
	\BibitemOpen
	\bibfield  {author} {\bibinfo {author} {\bibfnamefont {F.}~\bibnamefont
			{Ungar}}, \bibinfo {author} {\bibfnamefont {M.}~\bibnamefont {Cygorek}},\
		and\ \bibinfo {author} {\bibfnamefont {V.~M.}\ \bibnamefont {Axt}},\
	}\bibfield  {title} {\bibinfo {title} {Phonon-induced quantum ratchet in the
			exciton spin dynamics in diluted magnetic semiconductors in a magnetic
			field},\ }\href {https://doi.org/DOI:10.1103/PhysRevB.99.075301} {\bibfield
		{journal} {\bibinfo  {journal} {Phys. Rev. B}\ }\textbf {\bibinfo {volume}
			{99}},\ \bibinfo {pages} {075301} (\bibinfo {year} {2019})}\BibitemShut
	{NoStop}%
	\bibitem [{\citenamefont {Hamamoto}\ \emph {et~al.}(2019)\citenamefont
		{Hamamoto}, \citenamefont {Park}, \citenamefont {Ishizuka},\ and\
		\citenamefont {Nagaosa}}]{PRB2019scaling}%
	\BibitemOpen
	\bibfield  {author} {\bibinfo {author} {\bibfnamefont {K.}~\bibnamefont
			{Hamamoto}}, \bibinfo {author} {\bibfnamefont {T.}~\bibnamefont {Park}},
		\bibinfo {author} {\bibfnamefont {H.}~\bibnamefont {Ishizuka}},\ and\
		\bibinfo {author} {\bibfnamefont {N.}~\bibnamefont {Nagaosa}},\ }\bibfield
	{title} {\bibinfo {title} {Scaling theory of a quantum ratchet},\ }\href
	{https://doi.org/DOI:10.1103/PhysRevB.99.064307} {\bibfield  {journal}
		{\bibinfo  {journal} {Phys. Rev. B}\ }\textbf {\bibinfo {volume} {99}},\
		\bibinfo {pages} {064307} (\bibinfo {year} {2019})}\BibitemShut {NoStop}%
			\bibitem [{\citenamefont {Scheid}\ \emph {et~al.}(2007)\citenamefont {Scheid},
			\citenamefont {Pfund}, \citenamefont {Bercioux},\ and\ \citenamefont
			{Richter}}]{SR1}%
		\BibitemOpen
		\bibfield  {author} {\bibinfo {author} {\bibfnamefont {M.}~\bibnamefont
				{Scheid}}, \bibinfo {author} {\bibfnamefont {A.}~\bibnamefont {Pfund}},
			\bibinfo {author} {\bibfnamefont {D.}~\bibnamefont {Bercioux}},\ and\
			\bibinfo {author} {\bibfnamefont {K.}~\bibnamefont {Richter}},\ }\bibfield
		{title} {\bibinfo {title} {Coherent spin ratchets: A spin-orbit based quantum
				ratchet mechanism for spin-polarized currents in ballistic conductors},\
		}\href {https://doi.org/10.1103/PhysRevB.76.195303} {\bibfield  {journal}
			{\bibinfo  {journal} {Phys. Rev. B}\ }\textbf {\bibinfo {volume} {76}},\
			\bibinfo {pages} {195303} (\bibinfo {year} {2007})}\BibitemShut {NoStop}%
		\bibitem [{\citenamefont {Smirnov}\ \emph {et~al.}(2008)\citenamefont
			{Smirnov}, \citenamefont {Bercioux}, \citenamefont {Grifoni},\ and\
			\citenamefont {Richter}}]{DSR1}%
		\BibitemOpen
		\bibfield  {author} {\bibinfo {author} {\bibfnamefont {S.}~\bibnamefont
				{Smirnov}}, \bibinfo {author} {\bibfnamefont {D.}~\bibnamefont {Bercioux}},
			\bibinfo {author} {\bibfnamefont {M.}~\bibnamefont {Grifoni}},\ and\ \bibinfo
			{author} {\bibfnamefont {K.}~\bibnamefont {Richter}},\ }\bibfield  {title}
		{\bibinfo {title} {Interplay between quantum dissipation and an in-plane
				magnetic field in the spin ratchet effect},\ }\href
		{https://doi.org/10.1103/PhysRevB.78.245323} {\bibfield  {journal} {\bibinfo
				{journal} {Phys. Rev. B}\ }\textbf {\bibinfo {volume} {78}},\ \bibinfo
			{pages} {245323} (\bibinfo {year} {2008})}\BibitemShut {NoStop}%
		\bibitem [{\citenamefont {Smirnov}\ \emph {et~al.}(2009)\citenamefont
			{Smirnov}, \citenamefont {Bercioux}, \citenamefont {Grifoni},\ and\
			\citenamefont {Richter}}]{DSR2}%
		\BibitemOpen
		\bibfield  {author} {\bibinfo {author} {\bibfnamefont {S.}~\bibnamefont
				{Smirnov}}, \bibinfo {author} {\bibfnamefont {D.}~\bibnamefont {Bercioux}},
			\bibinfo {author} {\bibfnamefont {M.}~\bibnamefont {Grifoni}},\ and\ \bibinfo
			{author} {\bibfnamefont {K.}~\bibnamefont {Richter}},\ }\bibfield  {title}
		{\bibinfo {title} {Charge ratchet from spin flip: Space-time symmetry
				paradox},\ }\href {https://doi.org/10.1103/PhysRevB.80.201310} {\bibfield
			{journal} {\bibinfo  {journal} {Phys. Rev. B}\ }\textbf {\bibinfo {volume}
				{80}},\ \bibinfo {pages} {201310(R)} (\bibinfo {year} {2009})}\BibitemShut
		{NoStop}%
			\bibitem [{\citenamefont {Denisov}\ \emph {et~al.}(2007)\citenamefont
		{Denisov}, \citenamefont {Morales-Molina}, \citenamefont {Flach},\ and\
		\citenamefont {H\"anggi}}]{Denisov2007}%
	\BibitemOpen
	\bibfield  {author} {\bibinfo {author} {\bibfnamefont {S.}~\bibnamefont
			{Denisov}}, \bibinfo {author} {\bibfnamefont {L.}~\bibnamefont
			{Morales-Molina}}, \bibinfo {author} {\bibfnamefont {S.}~\bibnamefont
			{Flach}},\ and\ \bibinfo {author} {\bibfnamefont {P.}~\bibnamefont
			{H\"anggi}},\ }\bibfield  {title} {\bibinfo {title} {Periodically driven
			quantum ratchets: Symmetries and resonances},\ }\href@noop {} {\bibfield
		{journal} {\bibinfo  {journal} {Phys. Rev. A}\ }\textbf {\bibinfo {volume}
			{75}},\ \bibinfo {pages} {063424} (\bibinfo {year} {2007})}\BibitemShut
	{NoStop}%
	\bibitem [{\citenamefont {Salger}\ and\ \citenamefont
		{et~al.}(2009)}]{Science2009}%
	\BibitemOpen
	\bibfield  {author} {\bibinfo {author} {\bibfnamefont {T.}~\bibnamefont
			{Salger}}\ and\ \bibinfo {author} {\bibnamefont {et~al.}},\ }\bibfield
	{title} {\bibinfo {title} {Directed transport of atoms in a {H}amiltonian
			quantum ratchet},\ }\href {https://doi.org/DOI: 10.1126/science.1179546}
	{\bibfield  {journal} {\bibinfo  {journal} {Science}\ }\textbf {\bibinfo
			{volume} {326}},\ \bibinfo {pages} {1241} (\bibinfo {year}
		{2009})}\BibitemShut {NoStop}%
	\bibitem [{\citenamefont {Chen}\ and\ \citenamefont {el~al.}(2017)}]{QDK}%
	\BibitemOpen
	\bibfield  {author} {\bibinfo {author} {\bibfnamefont {L.}~\bibnamefont
			{Chen}}\ and\ \bibinfo {author} {\bibnamefont {el~al.}},\ }\bibfield  {title}
	{\bibinfo {title} {Quantum ratchet effect in a time non-uniform double-kicked
			model},\ }\href {https://doi.org/DOI:10.1142/S0217979217440635} {\bibfield
		{journal} {\bibinfo  {journal} {Int. J. Mod. Phys. B}\ }\textbf {\bibinfo
			{volume} {31}},\ \bibinfo {pages} {1744063} (\bibinfo {year}
		{2017})}\BibitemShut {NoStop}%
	\bibitem [{\citenamefont {Pellegrini}(2014)}]{Pellegrini}%
	\BibitemOpen
	\bibfield  {author} {\bibinfo {author} {\bibfnamefont {C.}~\bibnamefont
			{Pellegrini}},\ }\bibfield  {title} {\bibinfo {title} {Continuous time open
			quantum random walks and non-{M}arkovian {L}indblad master equations},\
	}\href {https://doi.org/DOI:10.1007/s10955-013-0910-x} {\bibfield  {journal}
		{\bibinfo  {journal} {J. Stat. Phys.}\ }\textbf {\bibinfo {volume} {154}},\
		\bibinfo {pages} {838} (\bibinfo {year} {2014})}\BibitemShut {NoStop}%
	\bibitem [{\citenamefont {Lee}\ and\ \citenamefont {Grier}(2005)}]{15}%
	\BibitemOpen
	\bibfield  {author} {\bibinfo {author} {\bibfnamefont {S.-H.}\ \bibnamefont
			{Lee}}\ and\ \bibinfo {author} {\bibfnamefont {D.~G.}\ \bibnamefont
			{Grier}},\ }\bibfield  {title} {\bibinfo {title} {Flux reversal in a
			two-state symmetric optical thermal ratchet},\ }\href@noop {} {\bibfield
		{journal} {\bibinfo  {journal} {Phys. Rev. E}\ }\textbf {\bibinfo {volume}
			{71}},\ \bibinfo {pages} {060102(R)} (\bibinfo {year} {2005})}\BibitemShut
	{NoStop}%
	\bibitem [{\citenamefont {Dupont}\ \emph {et~al.}(2023)\citenamefont {Dupont},
		\citenamefont {Gabardos}, \citenamefont {Arrouas}, \citenamefont {Ombredane},
		\citenamefont {Billy}, \citenamefont {Peaudecerf},\ and\ \citenamefont
		{Gu\'{e}ry-Odelin}}]{QR2023}%
	\BibitemOpen
	\bibfield  {author} {\bibinfo {author} {\bibfnamefont {N.}~\bibnamefont
			{Dupont}}, \bibinfo {author} {\bibfnamefont {L.}~\bibnamefont {Gabardos}},
		\bibinfo {author} {\bibfnamefont {F.}~\bibnamefont {Arrouas}}, \bibinfo
		{author} {\bibfnamefont {N.}~\bibnamefont {Ombredane}}, \bibinfo {author}
		{\bibfnamefont {J.}~\bibnamefont {Billy}}, \bibinfo {author} {\bibfnamefont
			{B.}~\bibnamefont {Peaudecerf}},\ and\ \bibinfo {author} {\bibfnamefont
			{D.}~\bibnamefont {Gu\'{e}ry-Odelin}},\ }\bibfield  {title} {\bibinfo {title}
		{Hamiltonian ratchet for matter-wave transport},\ }\href
	{https://doi.org/https://doi.org/10.1103/PhysRevLett.131.133401} {\bibfield
		{journal} {\bibinfo  {journal} {Phys. Rev. Lett.}\ }\textbf {\bibinfo
			{volume} {131}},\ \bibinfo {pages} {133401} (\bibinfo {year}
		{2023})}\BibitemShut {NoStop}%
	\bibitem [{\citenamefont {Valdez}\ \emph {et~al.}(2018)\citenamefont {Valdez},
		\citenamefont {Shchedrin}, \citenamefont {Heimsoth}, \citenamefont
		{Creffield}, \citenamefont {Sols},\ and\ \citenamefont {Carr}}]{PRL1202018}%
	\BibitemOpen
	\bibfield  {author} {\bibinfo {author} {\bibfnamefont {M.~A.}\ \bibnamefont
			{Valdez}}, \bibinfo {author} {\bibfnamefont {G.}~\bibnamefont {Shchedrin}},
		\bibinfo {author} {\bibfnamefont {M.}~\bibnamefont {Heimsoth}}, \bibinfo
		{author} {\bibfnamefont {C.~E.}\ \bibnamefont {Creffield}}, \bibinfo {author}
		{\bibfnamefont {F.}~\bibnamefont {Sols}},\ and\ \bibinfo {author}
		{\bibfnamefont {L.~D.}\ \bibnamefont {Carr}},\ }\bibfield  {title} {\bibinfo
		{title} {Many-body quantum chaos and entanglement in a quantum ratchet},\
	}\href {https://doi.org/DOI:10.1103/PhysRevLett.120.234101} {\bibfield
		{journal} {\bibinfo  {journal} {Phys. Rev. Lett.}\ }\textbf {\bibinfo
			{volume} {120}},\ \bibinfo {pages} {234101} (\bibinfo {year}
		{2018})}\BibitemShut {NoStop}%
			\bibitem [{\citenamefont {Paul}\ \emph {et~al.}(2023)\citenamefont {Paul},
			\citenamefont {B.},\ and\ \citenamefont
			{Kannan}}]{sanku}%
		\BibitemOpen
\bibfield  {author} {\bibinfo {author} {\bibfnamefont {S.}~\bibnamefont
		{Paul}}, \bibinfo {author} {\bibfnamefont {J. B.}~\bibnamefont {Kannan}},
	 \ and \bibinfo
	{author} {\bibfnamefont {M. S.}~\bibnamefont {Santhanam}},\ }\bibfield  {title} {\bibinfo {title}
	{Interaction-induced directed transport in quantum chaotic subsystems},\ }\href
{https://doi.org/https://doi.org/10.1103/PhysRevLett.131.133401} {\bibfield
	{journal} {\bibinfo  {journal} {Phys. Rev. E}\ }\textbf {\bibinfo
		{volume} {108}},\ \bibinfo {pages} {044208} (\bibinfo {year}
	{2023})}\BibitemShut {NoStop}%
	\bibitem [{\citenamefont {Zapata}\ \emph {et~al.}(1996)\citenamefont {Zapata},
		\citenamefont {Bartussek}, \citenamefont {Sols},\ and\ \citenamefont
		{H\"{a}nggi}}]{Zapata1996}%
		\BibitemOpen
	\bibfield  {author} {\bibinfo {author} {\bibfnamefont {I.}~\bibnamefont
			{Zapata}}, \bibinfo {author} {\bibfnamefont {R.}~\bibnamefont {Bartussek}},
		\bibinfo {author} {\bibfnamefont {F.}~\bibnamefont {Sols}},\ and\ \bibinfo
		{author} {\bibfnamefont {P.}~\bibnamefont {H\"{a}nggi}},\ }\bibfield  {title}
	{\bibinfo {title} {Voltage rectification by a {SQUID} ratchet},\ }\href
	{https://doi.org/10.1103/physrevlett.77.2292} {\bibfield  {journal} {\bibinfo
			{journal} {Phys. Rev. Lett.}\ }\textbf {\bibinfo {volume} {77}},\ \bibinfo
		{pages} {2292} (\bibinfo {year} {1996})}\BibitemShut {NoStop}%
	\bibitem [{\citenamefont {P. Faltermeier}\ and\ \citenamefont
		{et~al.}(2018)}]{PhysicaE2018}%
	\BibitemOpen
	\bibfield  {author} {\bibinfo {author} {\bibnamefont {P. Faltermeier}}\ and\
		\bibinfo {author} {\bibnamefont {et~al.}},\ }\bibfield  {title} {\bibinfo
		{title} {Circular and linear magnetic quantum ratchet effects in
			dual-grating-gate {C}d{T}e-based nanostructures},\ }\href
	{https://doi.org/https://doi.org/10.1016/j.physe.2018.04.001} {\bibfield
		{journal} {\bibinfo  {journal} {Phys. E: Low-Dimens. Syst. Nanostructures}\
		}\textbf {\bibinfo {volume} {101}},\ \bibinfo {pages} {178} (\bibinfo {year}
		{2018})}\BibitemShut {NoStop}%
	\bibitem [{\citenamefont {Drexler}\ and\ \citenamefont
		{et~al.}(2013)}]{NatureNano2013}%
	\BibitemOpen
	\bibfield  {author} {\bibinfo {author} {\bibfnamefont {C.}~\bibnamefont
			{Drexler}}\ and\ \bibinfo {author} {\bibnamefont {et~al.}},\ }\bibfield
	{title} {\bibinfo {title} {Magnetic quantum ratchet effect in graphene},\
	}\href {https://doi.org/DOI: 10.1038/NNANO.2012.231} {\bibfield  {journal}
		{\bibinfo  {journal} {Nat. Nanotechnol.}\ }\textbf {\bibinfo {volume} {8}},\
		\bibinfo {pages} {104} (\bibinfo {year} {2013})}\BibitemShut {NoStop}%
	\bibitem [{\citenamefont {Zhang}\ \emph {et~al.}(2015)\citenamefont {Zhang},
		\citenamefont {Li},\ and\ \citenamefont {Guo}}]{SciBull2015}%
	\BibitemOpen
	\bibfield  {author} {\bibinfo {author} {\bibfnamefont {C.}~\bibnamefont
			{Zhang}}, \bibinfo {author} {\bibfnamefont {C.-F.}\ \bibnamefont {Li}},\ and\
		\bibinfo {author} {\bibfnamefont {G.-C.}\ \bibnamefont {Guo}},\ }\bibfield
	{title} {\bibinfo {title} {Experimental demonstration of photonic quantum
			ratchet},\ }\href {https://doi.org/https://doi.org/10.1007/s11434-014-0710-y}
	{\bibfield  {journal} {\bibinfo  {journal} {Sci. Bull.}\ }\textbf {\bibinfo
			{volume} {60}},\ \bibinfo {pages} {2} (\bibinfo {year} {2015})}\BibitemShut
	{NoStop}%
	\bibitem [{\citenamefont {Cubero}\ \emph {et~al.}(2006)\citenamefont {Cubero},
		\citenamefont {Casado-Pascual}, \citenamefont {Morillo},\ and\ \citenamefont
		{H\"{a}nggi}}]{Polonica}%
	\BibitemOpen
	\bibfield  {author} {\bibinfo {author} {\bibfnamefont {D.}~
			{Cubero}}, \bibinfo {author} {\bibfnamefont {J.}~\bibnamefont
			{Casado-Pascual}}, \bibinfo {author} {\bibfnamefont {A.}~\bibnamefont
			{Alvarez}}, \bibinfo {author} {\bibnamefont {M.}~\bibnamefont
			{Morillo}},\ and\ \bibinfo {author} {\bibfnamefont {P.}~\bibnamefont
			{H\"{a}nggi}},\ }\bibfield  {title} {\bibinfo {title} {Overdamped
			deterministic ratchets driven by multifrequency forces},\ }\href@noop {}
	{\bibfield  {journal} {\bibinfo  {journal} {Acta Phys. Pol. B}\ }\textbf
		{\bibinfo {volume} {37}},\ \bibinfo {pages} {1467} (\bibinfo {year}
		{2006})}\BibitemShut {NoStop}%
	\bibitem [{\citenamefont {Budini}(2006)}]{budini2006}%
	\BibitemOpen
	\bibfield  {author} {\bibinfo {author} {\bibfnamefont {A.~A.}\ \bibnamefont
			{Budini}},\ }\bibfield  {title} {\bibinfo {title} {Lindblad rate equations},\
	}\href {https://doi.org/DOI: 10.1103/PhysRevA.74.053815} {\bibfield
		{journal} {\bibinfo  {journal} {Phys. Rev. A}\ }\textbf {\bibinfo {volume}
			{74}},\ \bibinfo {pages} {053815} (\bibinfo {year} {2006})}\BibitemShut
	{NoStop}%
		\bibitem [{\citenamefont {Derrida}(1983)}]{Derrida}%
	\BibitemOpen
	\bibfield  {author} {\bibinfo {author} {\bibfnamefont {B.}\ \bibnamefont
			{Derrida}},\ }\bibfield  {title} {\bibinfo {title} {Velocity and Diffusion Constant of a Periodic One-Dimensional Hopping Model},\
	}\href {https://doi.org/DOI: 10.1103/PhysRevA.74.053815} {\bibfield
		{journal} {\bibinfo  {journal} {J. Stat. Phys.}\ }\textbf {\bibinfo {volume}
			{31}},\ \bibinfo {pages} {433} (\bibinfo {year} {1983})}\BibitemShut
	{NoStop}%
	\bibitem [{\citenamefont {Cox}(1967)}]{Cox}%
	\BibitemOpen
	\bibfield  {author} {\bibinfo {author} {\bibfnamefont {D.~R.}\ \bibnamefont
			{Cox}},\ }\href@noop {} {\emph {\bibinfo {title} {Renewal Theory}}}\
	(\bibinfo  {publisher} {Methuen, London},\ \bibinfo {year}
	{1962})\BibitemShut {NoStop}%
	\bibitem [{\citenamefont {Breuer}\ and\ \citenamefont
		{Petruccione}(2007)}]{Breuer}%
	\BibitemOpen
	\bibfield  {author} {\bibinfo {author} {\bibfnamefont {H.~P.}\ \bibnamefont
			{Breuer}}\ and\ \bibinfo {author} {\bibfnamefont {F.}~\bibnamefont
			{Petruccione}},\ }\href@noop {} {\emph {\bibinfo {title} {The theory of open
				quantum systems}}}\ (\bibinfo  {publisher} {Oxford University Press, Oxford},\
	\bibinfo {year} {2007})\BibitemShut {NoStop}%
	\bibitem [{\citenamefont {van Kampen}(1979)}]{vanKampen}%
	\BibitemOpen
	\bibfield  {author} {\bibinfo {author} {\bibfnamefont {N.~G.}\ \bibnamefont
			{van Kampen}},\ }\bibfield  {title} {\bibinfo {title} {Composite stochastic
			processes},\ }\href {https://doi.org/DOI: 10.1103/PhysRevA.74.053815}
	{\bibfield  {journal} {\bibinfo  {journal} {Physica}\ }\textbf {\bibinfo
			{volume} {96A}},\ \bibinfo {pages} {435} (\bibinfo {year}
		{1979})}\BibitemShut {NoStop}%
		\bibitem [{\citenamefont {Henderson}\ \emph {et~al.}(2009)\citenamefont
			{Henderson}, \citenamefont {Ryu}, \citenamefont {MacCormick},\ and\
			\citenamefont {Boshier}}]{Optical1}%
		\BibitemOpen
		\bibfield  {author} {\bibinfo {author} {\bibfnamefont {K.}~\bibnamefont
				{Henderson}}, \bibinfo {author} {\bibfnamefont {C.}~\bibnamefont {Ryu}},
			\bibinfo {author} {\bibfnamefont {C.}~\bibnamefont {MacCormick}},\ and\
			\bibinfo {author} {\bibfnamefont {M.~G.}\ \bibnamefont {Boshier}},\
		}\bibfield  {title} {\bibinfo {title} {Experimental demonstration of painting
				arbitrary and dynamic potentials for Bose–Einstein condensates},\
		}\href@noop {} {\bibfield  {journal} {\bibinfo  {journal} {New J. 
					Phys.}\ }\textbf {\bibinfo {volume} {11}},\ \bibinfo {pages} {043030}
			(\bibinfo {year} {2009})}\BibitemShut {NoStop}%
		\bibitem [{\citenamefont {Gaunt}\ and\ \citenamefont
			{Hadzibabic}(2012)}]{Optical2}%
		\BibitemOpen
		\bibfield  {author} {\bibinfo {author} {\bibfnamefont {A.~L.}\ \bibnamefont
				{Gaunt}}\ and\ \bibinfo {author} {\bibfnamefont {Z.}~\bibnamefont
				{Hadzibabic}},\ }\bibfield  {title} {\bibinfo {title} {Robust digital
				holography for ultracold atom trapping},\ }\href@noop {} {\bibfield
			{journal} {\bibinfo  {journal} {Sci. Rep.}\ }\textbf {\bibinfo
				{volume} {2}},\ \bibinfo {pages} {721} (\bibinfo {year} {2012})}\BibitemShut {NoStop}%
			\bibitem [{\citenamefont {Shapira}\ and\ \citenamefont {Cohen}(2020)}]{QST}%
			\BibitemOpen
			\bibfield  {author} {\bibinfo {author} {\bibfnamefont {D.}~\bibnamefont
					{Shapira}}\ and\ \bibinfo {author} {\bibfnamefont {D.}~\bibnamefont
					{Cohen}},\ }\bibfield  {title} {\bibinfo {title} {Quantum stochastic
					transport along chains},\ }\href@noop {} {\bibfield  {journal} {\bibinfo
					{journal} {Sci. Rep.}\ }\textbf {\bibinfo {volume} {10}},\ \bibinfo {pages} {10353} (\bibinfo
				{year} {2020})}\BibitemShut {NoStop}%
		\bibitem [{\citenamefont {Windpassinger}\ and\ \citenamefont
			{Sengstock}(2013)}]{OL7}%
		\BibitemOpen
		\bibfield  {author} {\bibinfo {author} {\bibfnamefont {P.}~\bibnamefont
				{Windpassinger}}\ and\ \bibinfo {author} {\bibfnamefont {K.}~\bibnamefont
				{Sengstock}},\ }\bibfield  {title} {\bibinfo {title} {Engineering novel
				optical lattices},\ }\href@noop {} {\bibfield  {journal} {\bibinfo  {journal}
				{Rep. Prog. Phys.}\ }\textbf {\bibinfo {volume} {76}},\
			\bibinfo {pages} {086401} (\bibinfo {year} {2013})}\BibitemShut {NoStop}%
			\end{thebibliography}

\end{document}